\documentclass[]{aa}
\usepackage{graphicx}
\usepackage{longtable, lscape}
\usepackage{natbib}
\usepackage{amsmath}
\begin{document}
\title{Isolated elliptical galaxies and their globular cluster systems. II.  NGC 7796 - globular clusters, dynamics, companion 
 \thanks{Based
    on observations taken at the European Southern Observatory, Cerro Paranal, Chile,
    under the programme 089.B-0457. Partly based on observations taken at the Las Campanas Observatory and the Gemini observatory (GS-2011B-Q83).    }}
\subtitle{}
\titlerunning{NGC 7796}

\author{
T. Richtler 
 \inst{1,4} 
 \and
 R. Salinas
 \inst{2}
 \and
 R. R. Lane
 \inst{1}
 \and
 M. Hilker
 \inst{4}
 \and
 M. Schirmer
 \inst{3}
}
\offprints{T. Richtler}

\institute{
Departamento de Astronom\'{\i}a,
Universidad de Concepci\'on,
Concepci\'on, Chile;
tom@astro-udec.cl
\and
Department of Physics and Astronomy,
Michigan State University,
567 Wilson Road,
East Lansing, MI  48824;
\and
Gemini Observatory, Casilla 603, La Serena, Chile
\and
European Southern Observatory, Karl-Schwarzschild-Str.~2,
           D-85748 Garching, Germany 
}

\date{Received  / Accepted }

\abstract{ 
Rich globular cluster systems,
particularly the metal-poor part of them, are thought to be the visible manifestations of long-term accretion processes. The invisible part is the dark matter halo,
which may show some correspondence to the globular cluster system. It is therefore interesting to investigate the globular cluster systems of
isolated elliptical galaxies, which supposedly have not experienced  extended  accretion.}
{We investigate the globular cluster system of the isolated elliptical NGC 7796, present new photometry of the galaxy, and use published kinematical data to constrain
the dark matter content.}
{Deep images in B and R, obtained with the VIsible MultiObject Spectrograph (VIMOS) at the VLT, form the data base. We performed photometry with DAOPHOT and constructed a spherical
photometric model.   We present  isotropic and anisotropic Jeans-models and give a morphological description of the companion dwarf galaxy.   }
{The globular cluster system has about  2000 members, so it is not as rich as those of giant ellipticals in galaxy clusters with a comparable stellar mass, but richer than
many cluster systems of  other 
isolated ellipticals. 
The  colour distribution of GCs is bimodal,
which does not necessarily mean a metallicity bimodality. The kinematic literature data  are somewhat inconclusive. 
The velocity dispersion in the inner parts can be reproduced without dark matter under isotropy. Radially anisotropic models need a low stellar mass-to-light ratio, which would contrast with the
old age of the galaxy.  A MONDian model is supported by X-ray analysis and previous dynamical modelling, but better data are necessary for a confirmation.  The dwarf companion galaxy  NGC 7796-1 exhibits tidal tails, multiple nuclei, and very
boxy isophotes.
  }
{ NGC 7796 is an old, massive isolated elliptical galaxy with no indications of later major star formation events as seen frequently  in other isolated ellipticals. Its relatively rich globular cluster system shows that isolation does not always mean a poor cluster system.
The properties of the dwarf companion might indicate   a dwarf-dwarf merger.
 }
  
\maketitle

\section{Introduction}

The majority of elliptical galaxies are found in clusters and groups, while relatively
few are found in isolation. In spite of  their scarcity, isolated galaxies are plausibly  important test objects 
for scenarios of galactic  formation, given the relevance of the environment. 
Massive elliptical
galaxies assembled a significant part of their masses and sizes through
minor mergers during the past 10 Gyr \citep{vandokkum10}.  Galaxies in dense
environments are therefore expected to experience other modes of growth than
galaxies in low density regions. 
The early catalogues  of isolated galaxies \citep{karach73,karach97, turner76}, which
used angular distances from photographic plates as isolation criteria, 
 only contain  10\% that show an early-type morphology.  
 More modern morphological analyses based on SDSS-images identify about 3\% of  elliptical galaxies among the sample common
 to Karachentseva's catalogue and the SDSS \citep{hernandez08}, the majority showing some kind of distortion like diffuse haloes, shells, and
 dust lanes.  However, in another sample of nine extremely isolated galaxies, only two obvious merger remnants have been found \citep{marcum04}.
 
 Population studies of isolated ellipticals (IEs) have  been performed by \citet{kuntschner02} and \citet{collobert06}. They find typically younger ages and
 wider age spreads, as predicted from models of galaxy formation in
 a standard cosmological CDM scenario \citep{kauffmann96,niemi10}.   
Photometric properties of IEs  have been investigated by \citet{reda04}, who found that under closer scrutiny half of their galaxy sample  
show morphological peculiarities. \citet{smith04} presents an independent sample of IEs and studied their companions.
The work by \citet{hau06} is an important source for data regarding  the central  kinematics of IEs.
 
Many present morphologies with the signature of mergers that are
highly visible (see also  \citealt{tal09,lane13}). Others, even without obvious
visible disturbances, show intermediate ages through their dynamics. Given the
structural parameters,
the central
velocity dispersion is closely related to the stellar mass-to-light ratio (M/L), which in turn  can indicate
the age.  An
example is NGC 7507 with an $M/L$ value of 3.1 in the R-band, which is too low for an old
population \citep{salinas12}. 

A classical attribute of giant elliptical galaxies in clusters is their rich
globular cluster systems (GCSs).  There is now strong evidence that the majority
of metal-poor clusters (which dominate the GCSs of giant ellipticals) have
been accreted through the accretion of dwarf galaxies (see
\citealt{richtler13} for a recent review). One would, therefore, expect that
isolated ellipticals have poorer GCSs, and for most of the investigated
galaxies this is indeed the case \citep{spitler08,lane13,caso13}. 
However, the galaxy sample  of \citet{cho12} suggests that the role of the environment
is minor in comparison to the host galaxy mass as parameter.

  Within the $\Lambda$CDM framework,
it is reasonable to also expect  a connection of a hierarchically built-up  dark
matter halo with an accreted globular cluster system. For example, IEs should
have less dark matter than giant ellipticals in clusters \citep{niemi10}.

Little is known about the dark matter content of isolated
ellipticals, and what is known is sometimes not reliable. For example, NGC 7507 had the
reputation of a dark matter dominated galaxy \citep{krona00}, but a closer
look revealed that it is possible to construct models without dark matter
\citep{salinas12}.  
 
So it is of great interest to investigate more IEs 
 to find out whether NGC 7507 is just a weird outlier or whether it
represents a systematic still to be discovered.
Studies  centered on dark matter halos of isolated ellipticals  use in their majority X-ray data
 \citep{buote02,mulchaey99,osullivan07,memola11,milgrom12}. While the existence of dark matter
 halos (or, depending on the perspective, effects of modified gravity) are generally not questioned,
 the exact amount of dark matter in a given galaxy can be uncertain. However, NGC 7507 is not
 unique. Using the kinematics of planetary nebulae in NGC 4697, \citet{mendez09} found a declining
projected velocity dispersion, which leaves not much room for dark matter.
 
In this paper, we focus on NGC 7796, investigating its globular cluster
system, and developing a simple dynamical model on the basis of literature
data. NGC 7796 has been selected from the catalogue of IEs
by \citet{reda04}.  To our knowledge, this is the first publication
exclusively devoted to this galaxy.  NGC 7796 has no immediate neighbours with
known radial velocities, but, as we show in this paper, has a dwarf companion
with a tidal tail. Table \ref{tab:basicdata} lists the basic data.

The NED\footnote{NASA Extragalactic Database;http://ned.ipac.caltech.edu/}
lists seven dwarf galaxies with similar redshifts within a projected distance
of 400-600 kpc.  Spectroscopy of the central regions indicates that the galaxy
is old and metal-rich \citep{bertin94,thomas05,milone07,beuing02}.  NGC 7796
is among the galaxies dynamically modelled by \citet{magorrian01} (based on the Bertin et al. 1994 data), 
who found
evidence for a dark halo, and it has a counterrotating core \citep{bettoni01}.

\begin{table}[]
\caption{Basic and derived parameters for NGC 7796}
\begin{center}
\begin{tabular}{lll}

Parameter   &   Value  &  Ref. \\
 \hline
Right Ascension (2000) &  23h 58m 59.8s    & NED          \\  
Declination    (2000)      &   -55d 27m 30s   & NED           \\
rad. velocity     & 3364 km/s  & NED \\
distance & 49.9 Mpc & \citet{tonry01}  \\
scale &   1\arcsec\ = 242.4 pc &    \\
$R_{eff}$   &   21\arcsec\  &   RC3\\
$R_{eff}$  &  31.8\arcsec\ & present paper ($<$300\arcsec\ )\\
$V_T$    &   11.49    &   NED  \\
$A_V$   &  0.03 &    NED \\
B-V &   1.00    &                 \citet{poulain88} \\
U-B  &    0.70  &               "                   \\
V-R  & 0.56     &                "                   \\
\hline
\end{tabular}
\end{center}
\label{tab:basicdata}
\end{table}%

We adopt a distance of 50 Mpc, corresponding to 242.4 pc/arcsec.  This  paper is the second in a series on isolated
ellipticals. Paper I is \citet{lane13}.

\section{Observations and reductions}

\subsection{Very Large Telescope/VIMOS}
\begin{figure}[]
\begin{center}
\includegraphics[width=0.4\textwidth]{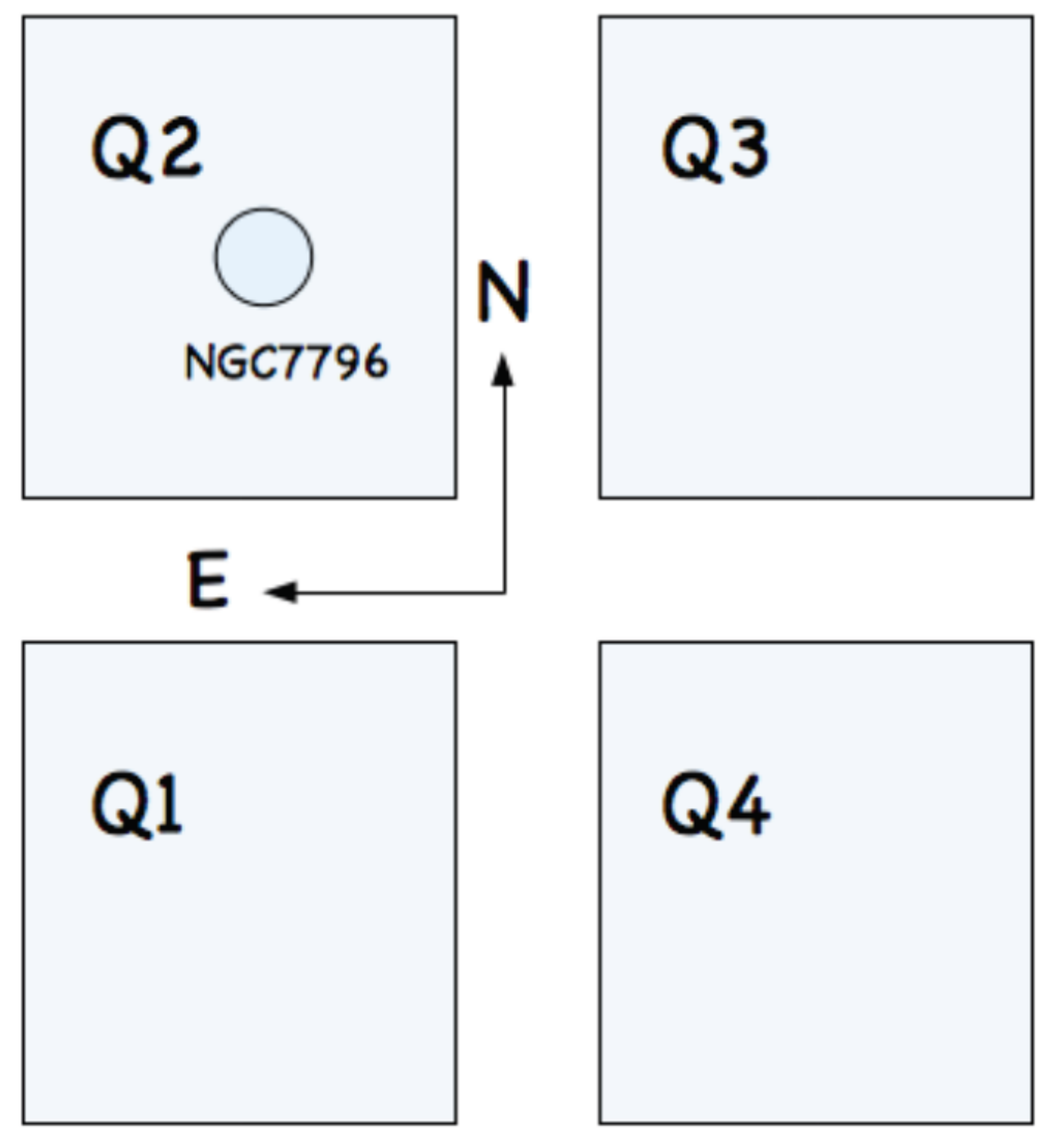}
\caption{ Arrangements of the VIMOS quadrants for a position angle of 90$^\circ$. The chips have sizes of 7\arcmin\ $\times$ 8\arcmin\ . The chip gaps have a width of 2\arcmin\ . The VIMOS field in the distance of NGC 7796 therefore covers 235$\times$209 kpc$^2$. NGC 7796 is centered on quadrant 2.
}
\label{fig:quadrants}
\end{center}
\end{figure}

The observations were taken using the ESO VLT/UT3 (\textit{Melipal})/VIsible
MultiObject Spectrograph (VIMOS) on the nights of July 22 and 24, 2012. VIMOS
has a nominal field of view of $4\times7\arcmin\times 8\arcmin$ separated by
2\arcmin~gaps, with a pixel scale of 0.205\arcsec~(see Fig
\ref{fig:quadrants}). Images were obtained through Bessell $B$ and $R$ filters,
with total exposure times of $8\times830$ and $16\times430$ seconds,
respectively.

Image reduction was done with the reduction recipes within the ESO VIMOS
pipeline (v.2.6.14) operated through \textsc{gasgano} (v.2.4.3). Master bias
and flat frames were produced from calibrations taken on the same nights as
the science observations. Bias removal, flat fielding and median combination
for each VIMOS quadrant were subsequently done with the recipe
\emph{vmimobsjitter}. The final combined images have seeing values of
$0.7\arcsec$ and $0.74\arcsec$ in $B$ and $R$, respectively, averaged over the
four quadrants.

\subsection{Magellan/Baade/IMACS}
The idea behind our IMACS observations was to image a large field with deep
photometry and to search for tidal features on a larger scale than possible
with VIMOS (see Sec.\ref{sec:imacs}). We observed NGC 7796 with the focal
reducer IMACS at the 6.5m Magellan/Baade telecope at the Carnegie observatory
at Las Campanas, Chile, on the night of August 8, 2013. In the f/2 mode, IMACS
provides a circular field of 27\arcmin\ with a pixel scale of
0.2\arcsec/pixel. Eleven 600 sec exposures were taken in Bessell V2 with two
overlapping pointings. The seeing on the final image is 0.76\arcsec.

The removal of instrumental signatures, astrometric alignment, and co-addition
of the 11 individual images was carried out using the THELI-data reduction
pipeline \citep{erben05,schirmer13}.

\section{Photometry with the VIMOS data}

The photometry was performed on images with the galaxy light removed.  To
model the galaxy light, we applied a quadratic median-filter with a
side-length of 15 pixels. Given the FWHM of the point-spread-function (PSF) of
3.5 pixels, the median size is large enough to not modify point-sources.

For the photometry of GC candidates, we used DAOPHOT within the MIDAS
environment. We chose the parameters of the finding routine to find about
10\,000 objects in each quadrant.  The actual number of remaining sources at
faint magnitudes is then determined by the PSF-photometry, which discards many
objects, and the matching between the $B$- and the $R$-filter.  Typically 15
PSF-stars in each quadrant were found. The {\it allstar}-routine selected
about 2/3 of all sources. Matching the lists in the $B$ and $R$ filters, many
of the faintest objects did not find partners, which had been searched for
with a maximum displacement of 5 pixels. Finally, the numbers of objects with
both $B$ and $R$ photometry were the following: Q1: 5222; Q2:5742; Q3:5299;
Q4:6717.

\subsection{Photometric calibration} 
\label{sec:calibration}

The photometry was calibrated into the standard Johnson B and R
with observations of the \citet{landolt92} field
PG1323 which also includes several secondary Stetson standards \footnote{

http://www3.cadc-ccda.hia-iha.nrc-cnrc.gc.ca/community/STETSON/standards/}.

The transformation equations to the standard system used were
\begin{equation}
\label{eq:calibration1}
B_{\rm obs} = B_{\rm std} + Z_B + K_B (X-1) + b_B (B-R)_{\rm std}
\end{equation}
\begin{equation}\label{eq:calibration2}
R_{\rm obs} = R_{\rm std} + Z_R + K_R (X-1) + b_R (B-R)_{\rm std}
\end{equation}

where $X$ is the airmass of the observation. Coefficients were obtained using
\textsc{iraf/photcal}. The observations have been done in photometric conditions, but 
the airmass coverage was too small for an independent calibration, so the extinction coefficients were fixed to values
$K_B=0.242$ and $K_R=0.08$ taken from the ESO webpages. The rest of the
coefficients for each quadrant can be seen in Table \ref{table:calibration}.

\begin{table*}
  \caption{Transformation equation coefficients for Eqs.
\ref{eq:calibration1}
    and \ref{eq:calibration2} for the different quadrants.}
\label{table:calibration}
\centering
 \begin{tabular}{@{}lccccccc@{}}
Quadrant    &   \multicolumn{2}{c}{Zero-point} &
\multicolumn{2}{c}{Color term} &N$_{\rm stars}$ &
\multicolumn{2}{c}{rms}\\
& $B$ & $R$ & $B$ & $R$& &$B$&$R$\\
\hline
Q1 & $27.477 \pm 0.012$& $27.371 \pm 0.029 $& $-0.015 \pm 0.011$ &$
0.030 \pm 0.006$ & 24 & 0.034 & 0.023\\
Q2 & $27.440 \pm 0.013$& $27.361 \pm 0.008 $& $ 0.010 \pm 0.010$ &$
0.008 \pm 0.006$ & 29 & 0.026 & 0.022\\
Q3 & $27.421 \pm 0.028$& $27.197 \pm 0.017 $& $ 0.041 \pm 0.026$ &
$-0.025 \pm 0.014$ & 24 & 0.068 & 0.035\\
Q4 & $27.560 \pm 0.011$& $27.326 \pm 0.016 $& $ 0.011 \pm 0.017$ &
$-0.018 \pm 0.014$ & 20 & 0.042 & 0.038\\
\hline
\end{tabular}
\end{table*}

\subsection{Comparison with aperture photometry}
\begin{figure}[]
\begin{center}
\includegraphics[width=0.4\textwidth]{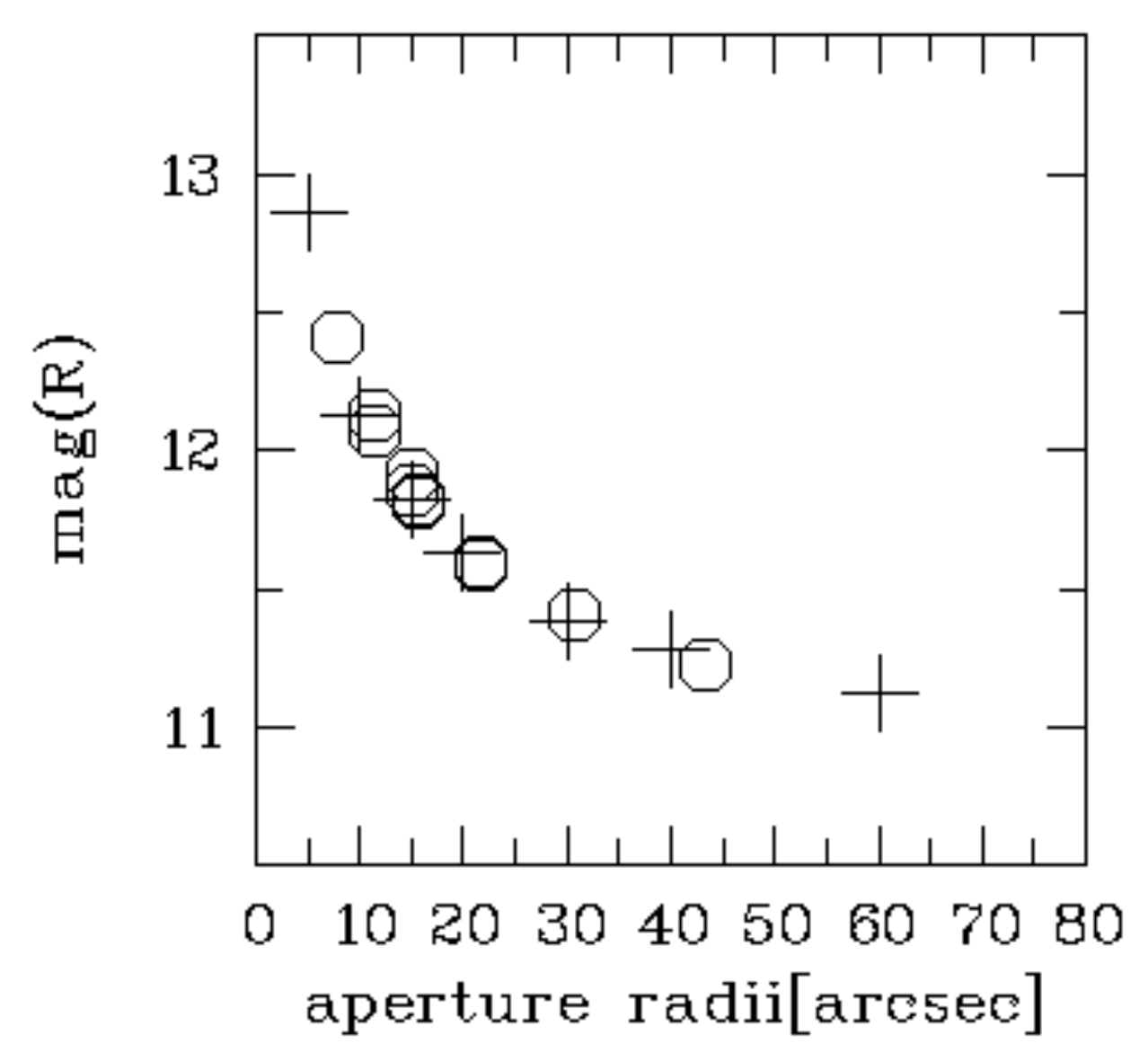}
\caption{Check of our
  R-band calibration. Polygons: Aperture photometry from
  \citet{prugniel98}. Crosses: radii of simulated apertures from our model profile.}
\label{fig:apertures}
\end{center}
\end{figure}

Support for the quality of our calibration comes from the comparison with
aperture photometry. We use our spherical light model (described in section
\ref{sec:profile}) to simulate aperture photometry by
\citealt{prugniel98}). This is shown in Fig.\ref{fig:apertures}, where the
simulated apertures are compared with measurements. The agreement is
excellent, to within 0.02 mag.

\subsection{Point source selection}

 \begin{figure}[h!]
\begin{center}
\includegraphics[width=0.4\textwidth]{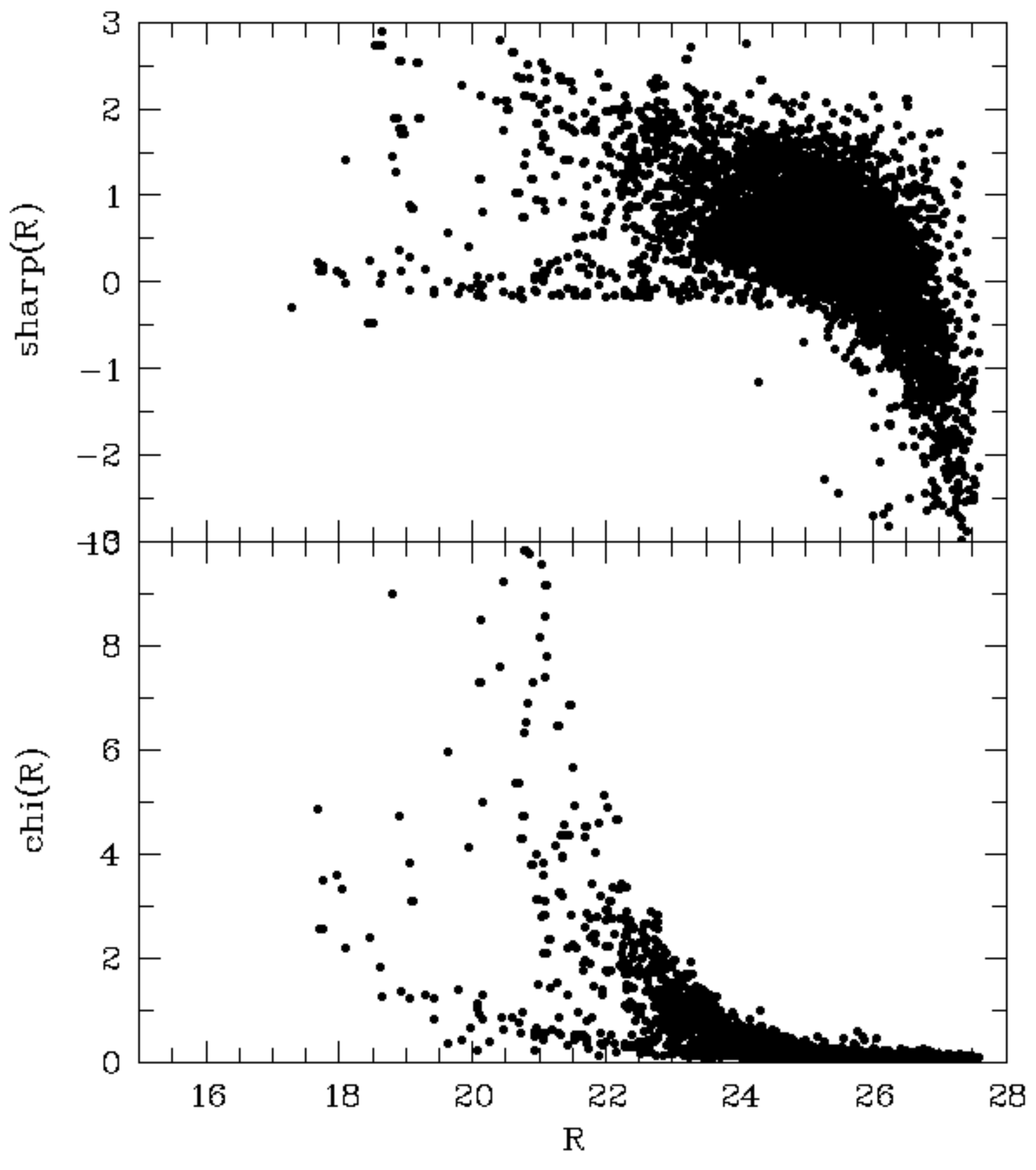}
\caption{The distribution of the DAOPHOT $chi$ and $sharp$ parameters is shown
  for Quadrant 4 in the $R$-band. The different sequences of stellar-like
  objects and galaxies are easily distinguished. We apply the selection {\it
    chi} less than 1 and {\it sharp} between -0.5 and 1\label{fig:chisharp}.}
\end{center}
\end{figure}

We did not apply very strict selection criteria to maintain the depth of the
photometry. Fig.\ref{fig:chisharp} shows as an example the distribution of
{\it chi} and {\it sharp}-values for Quadrant 4 in the $R$-band.  The DAOPHOT
{\it chi}-parameter measures the chi-square deviation, when fitting the PSF to
given objects, and the {\it sharp}- parameter measures kurtosis-like
deviations (see the DAOPHOT users manual for more information). Ideally, both
parameters should be zero for stars and ideal PSFs. The two sequences for
stellar-like objects and galaxies are discernible. For magnitudes fainter than
about R=24 mag, stellar-like and extended objects partly merge regarding their
parameters and a strong restriction only results in a pronounced
incompleteness at faint magnitudes. We therefore apply only a mild selection
using the R-band photometry.  The selection is: {\it chi} less than 1 and {\it
  sharp} between -0.5 and 1. This leaves 3728, 3171, 3351, and 4548 objects in
Q1-Q4, respectively. The photometry in Q4 clearly is the deepest (see Fig.\ref{fig:CMDs}).

\section{The globular cluster system}

\subsection{Colour-magnitude diagram}
\begin{figure*}[th!]
\begin{center}
\includegraphics[width=0.9\textwidth]{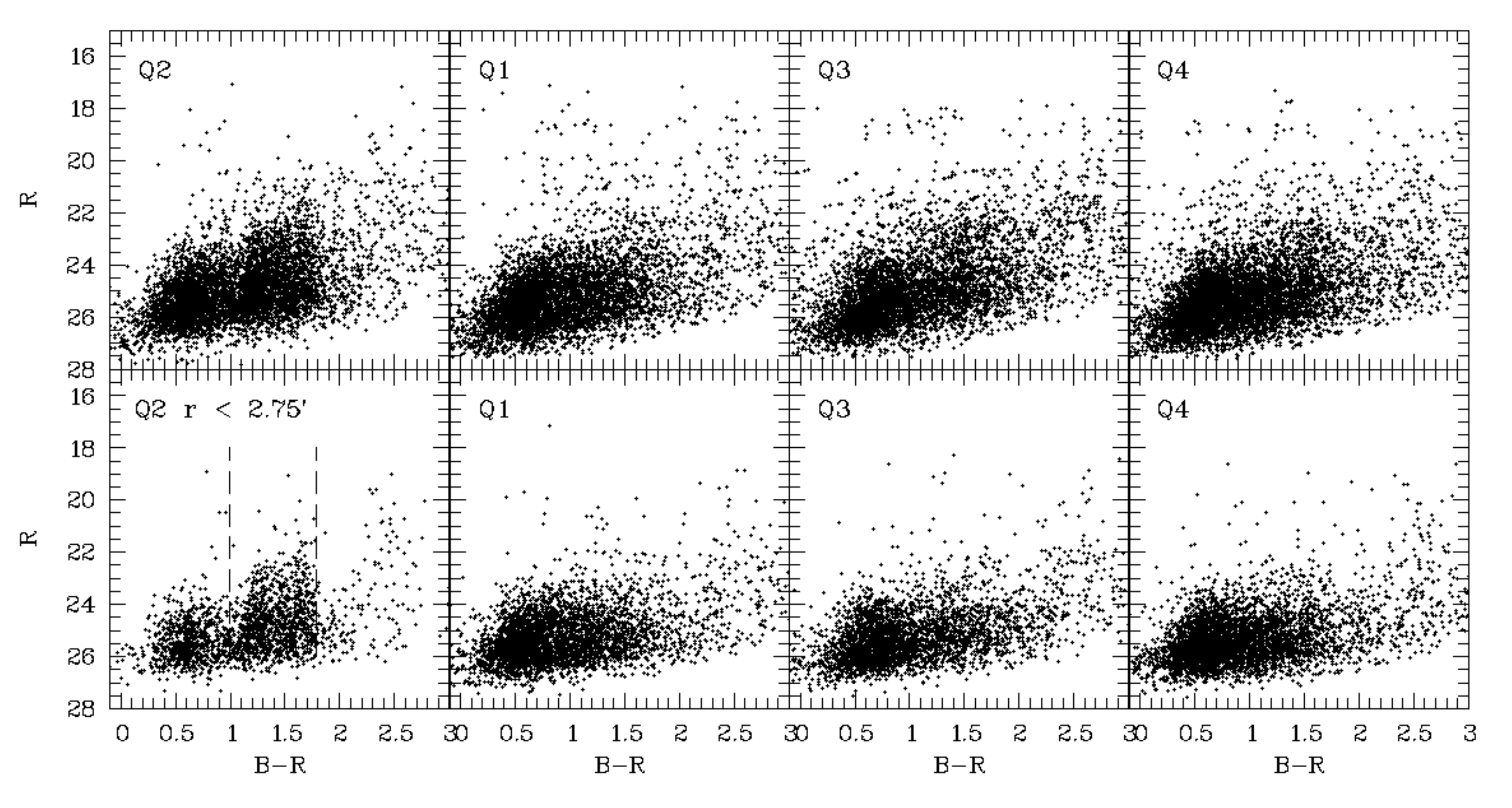}
\caption{The upper panels show the CMDs of the single quadrants without any
  selection. The lower row uses the point source selection as described in the
  text. Moreover, the Q2-CMD plots only objects with distances less than
  2.75\arcmin\ to the galaxy center. The GCS of NGC 7796 is very prominently
  seen, including its bimodal colour distribution. Note the sharp borders in
  colour. The vertical dashed lines mark the metallicities Z=0.001 ([Fe/H]=$-2.3$)
  and Z=0.04 ([Fe/H]=+0.3) for GCs of 11.2 Gyr
  according to \citet{marigo08}. Some of the brighter objects of the ($R=22$
  mag corresponds to $M_R = -11.5$) must be bright globular clusters as well.}
\label{fig:CMDs}
\end{center}
\end{figure*}
The arrangement of the VIMOS chips are shown in Fig.\ref{fig:quadrants}. The
projected distance of the frame border of Q3 is about 75 kpc away from the
centre of NGC 7796. The parts of quadrants 3 and 1, which lie nearest to NGC
7796, may therefore contain still some GCs.  The foreground reddening towards
NGC 7796 is small (E[B-R]=0.015 mag according to \citealt{schlegel98}) and we
neglect it.  The upper panels in Fig.\ref{fig:CMDs} show the CMDs in the
various quadrants without applying any selections.  Quadrant 2 contains NGC
7796 and its entire field is plotted.  The GC system is immediately striking.
The lower panels show the CMDs with the previously discussed selection for
point sources.  Moreover, a radius selection is applied for Q2, and the GCS
appears even more prominent, and a bimodality is visible without any
statistical aid. The vertical dashed lines indicate the theoretical colour of
an old GC population for the metallicities [Fe/H]=$-$2.3 and solar metallicity,
using the models of \citet{marigo08}. The red (metal-rich) limit is precisely
defined. This gives weight to the observation that the blue (metal-poor in the
case of old clusters) part is more or less empty in the regime of bright
clusters.

The vertical sequence of blue clusters seems to be tilted.  We do not think
that this is a metallicity effect and we return to this point in the
discussion.

\subsection{Colour distribution}

\begin{figure*}[]
\begin{center}
\includegraphics[width=0.7\textwidth]{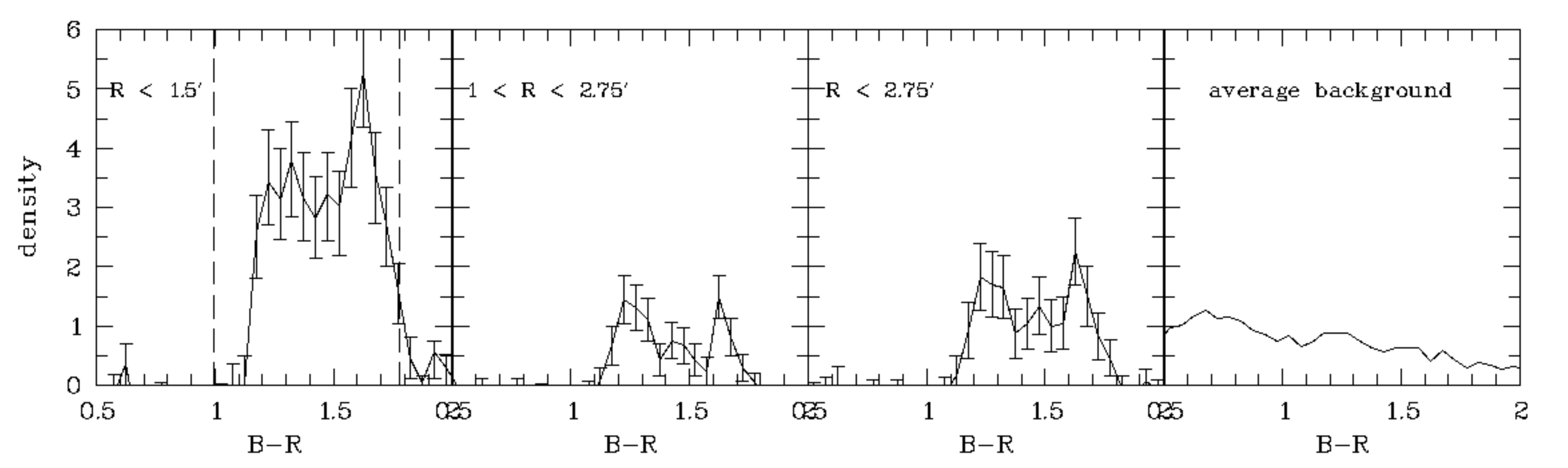}
\caption{Surface density of GCs vs. the colour for three different
  radial bins. The vertical dashed lines in the left panel indicate the colour of the models of
\citet{marigo08} for 11.2 Gyr old clusters and for the metallicities Z=0.001
      ([Fe/H]=$-$2.3; $B-R=0.99$) and Z=0.04 ([Fe/H]=+0.3; $B-R=1.78$).
  The rightmost panel is the background. The distinct bimodal appearance remains visible at all
  radii. 
  }
\label{fig:colourhist}
\end{center}
\end{figure*}

We evaluate the colour distribution in bins of 0.05 mag, using the quadrants
Q1,Q3,Q4 as background fields to subtract the background from
Quadrant 2.  The limiting magnitude was set to 25.5 to avoid the faintest
luminosity bins, where the photometry in the region of NGC 7796 is strikingly  less complete than in the other
quadrants.
Fig.\ref{fig:colourhist} shows the results, with surface densities (numbers
per square arc min) plotted against colour.

The leftmost panel shows the cluster system within 1.5\arcmin\ (21.8 kpc),
where the red clusters slightly dominate.  The middle panel depicts the region
between 1\arcmin\ and 2.75\arcmin\ (14.5 kpc and 40 kpc).  The right panel shows
the entire cluster system within 2.75\arcmin.

The background subtraction leaves a clean colour interval $1.1 < B-R < 1.75$
corresponding to almost the full metallicity spread of old globular clusters.
The vertical dashed lines in Fig.\ref{fig:colourhist}  indicate the colour of the models by
\citet{marigo08} for 11.2 Gyr old clusters and for the metallicities
      [Fe/H]=$-$2.3 ($B-R=0.96$) and [Fe/H]=0.04 ($B-R=1.78$).  Younger GCs in
      larger numbers, which are bluer than $B-R=0.95$, are absent.  
The relative amplitudes of the blue and the red peaks do not change much with
radius, indicating that the respective density profiles are not very
different.  The inner red peak may be a bit higher than the blue peak, but
only marginally, given the uncertainties. At least, it is very different from,
for example, Fig.10 by \citet{dirsch03}, where the red GC population in NGC
1399 becomes minor at large distances. The shallower profile of the blue GCs,
which is quite common in rich GCSs, probably reflects the assembly process by
the infall of dwarf galaxies.  In Sec.\ref{sec:density} we directly compare
the number density profiles of blue and red clusters with the galaxy light
profile and do not see striking differences.


The widely discussed bimodality (e.g. \citealt{richtler13}), although details
are still under debate, has a quite well accepted explanation. We make
respective remarks in the discussion.

\subsection{Number counts and specific frequency}

To estimate the total number of clusters, one should ideally know the
luminosity function (LF) down to $1-2$ mag below the turn-over magnitude
(TOM). The absolute TOM in the $R$-filter in {\it old} GCSs is about $-7.9$
\citep{richtler03,reijkuba12}.  We therefore expect the TOM to be around
$R=25.6$, whereas our photometry is already incomplete at brighter
magnitudes. One possibility would be to perform experiments with artificial
stars to determine completeness factors in dependence on radius and
magnitude. This would be a quite complex procedure, because the completeness
in the different quadrants probably depends in a complicated way on the
matching procedure, and it remains to be shown whether the completeness of
artificial stars is the same as for real star
images.
 Moreover, because we can renounce a precise
knowledge of the completeness factors, we prefer an easier method.

\begin{figure}[h]
\begin{center}
\includegraphics[width=0.5\textwidth]{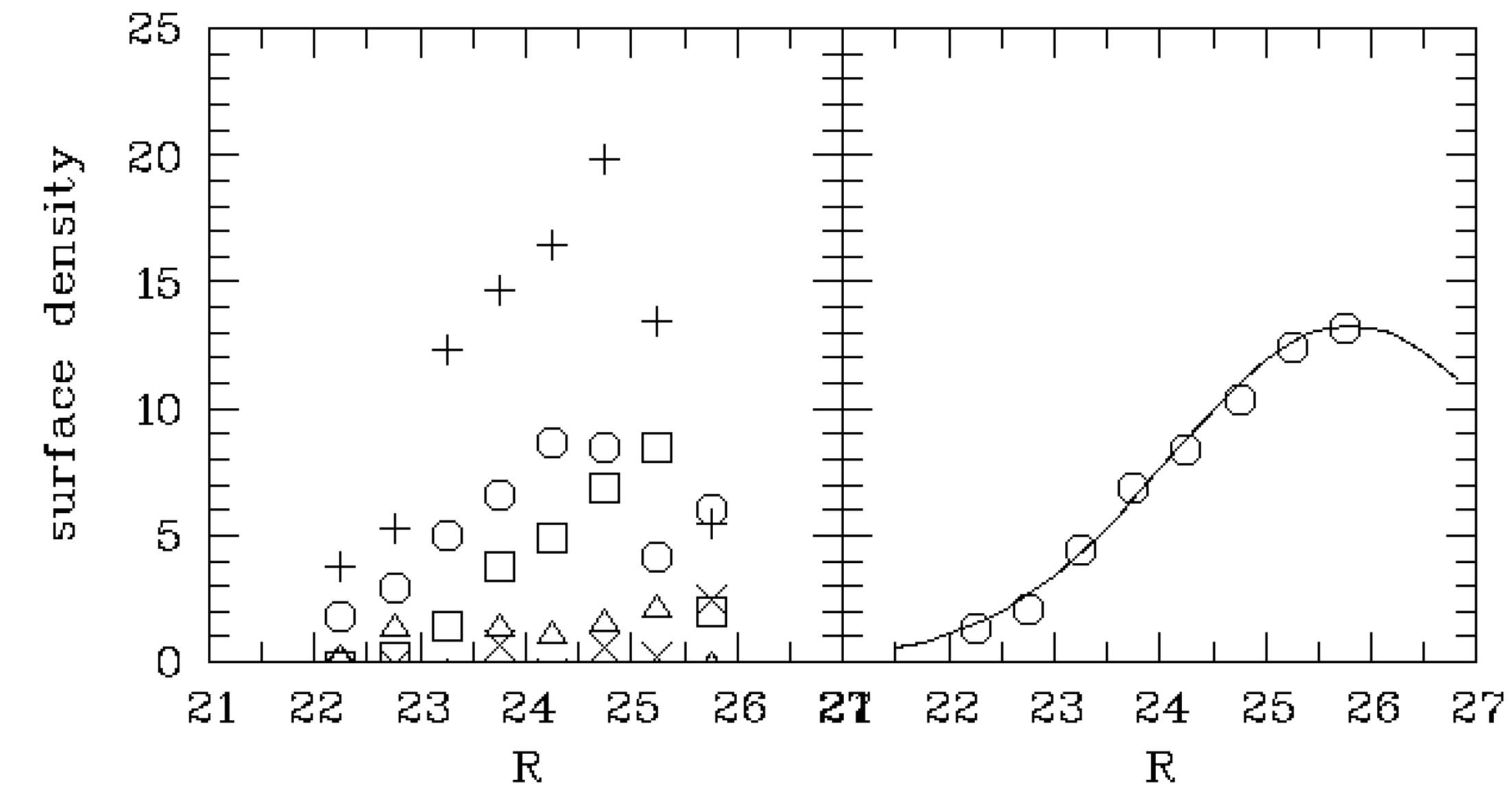}
\caption{Left panel:  LFs of clusters (numbers per square arcmin) for different radius
  intervals. Under the assumption that the LFs are not radius dependent, we
  estimate the incompleteness in the two faintest magnitude bins. (1) Plus signs:
  $0.15\arcmin-0.8\arcmin$; (2) polygons: $0.8\arcmin-1.3\arcmin$; (3) squares:
  $1.3\arcmin-1.8\arcmin$; (4) triangles: $1.8\arcmin-2.3\arcmin$; (5) crosses:
  $2.3\arcmin-2.8\arcmin$. Right panel: average density after corrections in the radial
  interval $0.15\arcmin - 1.8\arcmin$. A Gaussian with a peak magnitude of R=25.8 and $\sigma_R$=1.7 is a good
  fit to the data.}
\label{fig:completeness}
\end{center}
\end{figure}

In Fig.\ref{fig:completeness}, left panel, the surface densities (background
subtracted,   see Table \ref{tab:LF} for the background counts) of GCs in 5 different radius intervals are plotted. It is clear
that the completeness factors are quite different. The inner region offers
good statistics, but the galaxy light makes the faintest magnitude bins
incomplete.  The outer bins are more complete, but the number density becomes
so low that the uncertainty of the background subtraction becomes dominant.
As a compromise between number statistics and completeness, we adopt the
LF No.3 (see caption Fig.\ref{fig:completeness})
 (squares) as reference, assuming that it is
complete until $R=25.25$ and that the last bin is the TOM. Therefore, we
adjust the last bin by shifting it to the same density of the bin at R=25.25, that means a value of 8.5.
  We
do the same with the two inner LFs. The adopted densities are given in Table
\ref{tab:LF} in parentheses. The two outer LFs are not longer
considered because of their low numbers.  The right panel of Fig.\ref{fig:completeness} shows the total
``corrected'' densities. The resulting total number of objects brighter than the TOM  according to Table \ref{tab:LF} within a
projected radial annulus  of 0.15\arcmin\ - 1.8\arcmin\ (corresponding to 2.2 kpc  - 26.2 kpc) is then 589$\pm$47.
 
The uncertainties are calculated by assuming Poisson errors for the original count rates. The background has been determined as the average background
in the quadrants Q1,Q3,Q4 with the areas 46.23, 41.1, and 47.72 square arcmin, respectively. 

If we assume that the number density profile of GCs in NGC 7796 follows the galaxy
light (we justify that with Fig.\ref{fig:density}), we can use our photometric model to reproduce the inner GC densities
and calculate the fraction of clusters, which fall outside this range.
We (somewhat arbitrarily, but introducing only  a small error)
cut the GCS at 10 effective radii (77 kpc). The luminosities in the three areas $<$0.15\arcmin\ , 0.15\arcmin\ - 1.8\arcmin\ , 1.8\arcmin\ - 5.35\arcmin\ 
are 1.96$\times10^{10} L_\odot$, 4.93$\times10^{10} L_\odot$, and 1.97$\times10^{10} L_\odot$, respectively.
In proportion to the luminosities, we assign to the inner area 0.4$\times$589 GCs, to the outer area the same number
and get 1060 GCs. Since that is the number until the TOM, we double it and get 2120$\pm$170 GCs as the grand total, scaling the uncertainty of the counts with the
total number.
 The absolute $R$-magnitude in the same
area is $-22.9$. With $V-R = 0.6$, one gets $S_N=2.5$ for the specific
frequency. 
To estimate the final uncertainty, we allow 
10\% uncertainty in the distance, which means an uncertainty of 0.22 mag of the absolute magnitude.  The squared uncertainties in $S_N$ from the
distance and counts are  then 0.22  and 0.04, respectively, and the final error is 0.5.

Values of other IEs are lower (e.g. \citealt{caso13}), but  intermediate-age and therefore brighter
stellar populations contribute significantly to a lower $S_N$. 
 
 The globular cluster luminosity function (GCLF) can serve as a good distance indicator \citep{richtler03,reijkuba12,villegas10}
 provided that the  GCLF reaches beyond the turn-over magnitude (TOM). This is not the case with the present data,
 but in the right panel of Fig.\ref{fig:completeness} we show that a Gaussian with $\sigma_R$=1.7 and a TOM of R=25.8
 is a good representation and fits to the adopted distance of 50 Mpc.   Since the TOM is barely reached, we cannot say anything
 about a possible population of younger clusters.

\begin{table*}[ht!]
\caption{Background subtracted counts in radial and magnitude bins. Values are
  given in units of objects/square arcmin. The total number is the (corrected)  density multiplied
  with the area.}
 \label{tab:LF} 
\begin{center}
\begin{tabular}{lllllll}
 mag/radius    & 0.15-0.8 arcmin  &0.8-1.3 arcmin   &1.3-1.8 arcmin    & 1.8-2.3 arcmin& 2.3-2.75 arcmin &  background  \\
\hline
 22.25 &  3.8$\pm$1.5           & 1.8 $\pm$0.8                            & 0$\pm$0.2              & 0.1$\pm$0.3              & 0 &     0.36$\pm$  0.05 \\
 22.75 &  5.2  $\pm$1.7        & 3.0 $\pm$1.0                              & 0.2$\pm$0.4            & 1.3$\pm$0.5              & 0  &    0.37$\pm$0.05   \\
 23.25 &  12.3$\pm$2.6       & 5.0  $\pm$1.4                             &   1.4$\pm$0.7           &   0$\pm$0.3             &  0  &      1.07$\pm$  0.09\\ 
 23.75 & 14.7$\pm$2.9       & 6.6  $\pm$1.6                                &  3.8$\pm$1.0          &  1.3$\pm$0.7            & 0.6  &   1.32$\pm$0.10\\    
 24.25 & 16.4$\pm$3.2         &  8.7$\pm$1.9                                 &  4.9$\pm$1.3          &  1.0$\pm$0.8            &   0  & 3.15$\pm$0.15 \\
 24.75 &  19.9$\pm$3.6        & 8.5$\pm$2.0 (10$\pm$2.3)              & 6.8$\pm$1.6        & 1.5$\pm$1.0           & 0.6  & 4.87$\pm$0.19\\
 25.25 & 13.5 $\pm$3.2 (23$\pm$ 5.4 )   &  4.2$\pm$1.8 (12$\pm$5.1)   &  8.5$\pm$1.8  &  2.1$\pm$1.2   &  0.2   &   6.10$\pm$0.21 \\
 25.75  & 5.4  $\pm$2.5 (23$\pm$ 10.6)    & 6.0$\pm$2.0 (12$\pm$4.0)  &   2.0 $\pm$1.4(8.5$\pm5.9)$     &  0$\pm$  1.0    &  2.4 & 6.40$\pm$0.22  \\
  total   &   229$\pm$26  &  194$\pm$25       &     166$\pm$30    &  -   & - \\
 \hline
\end{tabular}

\end{center}
\label{tab:completeness}
\end{table*}%

\subsection{Number density profile}
\label{sec:density}
The density profile of a GCS permits a qualitative assessment of the relation
between the GCS and its host galaxy. For example, if the metal-poor population of a GCS
has a shallower density profile than its metal-rich counterpart or its host
galaxy, as it is normally the case, then it may be interpreted as the effect
of infall processes which are responsible for a large part of the GC
population.

\begin{figure}[]
\begin{center}
\includegraphics[width=0.5\textwidth]{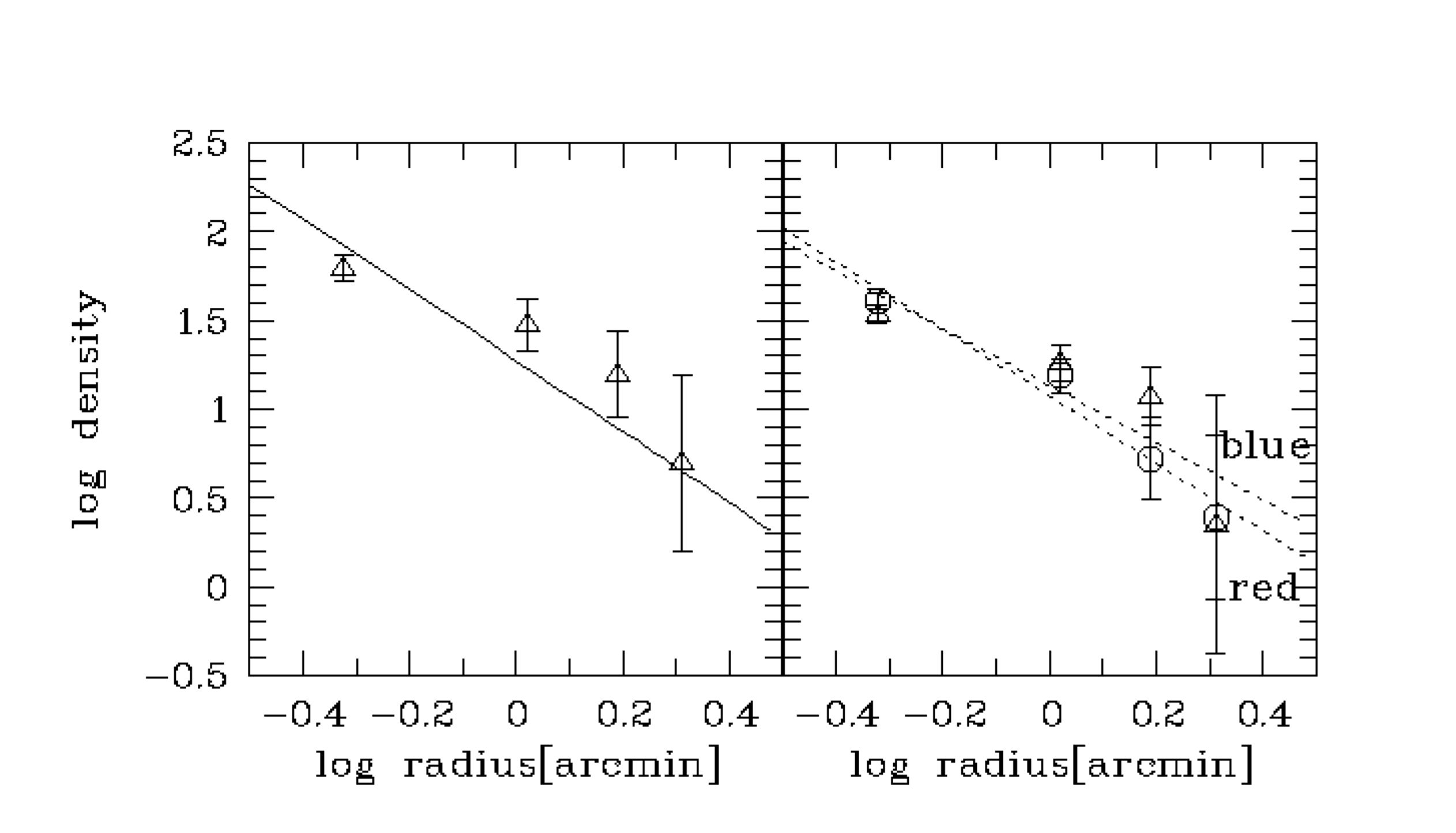}
\caption{Left panel: density profile (logarithm of numbers per square
  arcmin) of all GC candidates brighter than $R=25$ mag. The straight line is
  the arbitrarily scaled light profile of the galaxy for showing that these
  profiles are not distinguishable. 
    Right panel: the density profiles for red (circles) and blue (triangles) clusters
  separately.  The labels ''blue'' and  ''red'' indicate the respective linear fits. The blue clusters show a marginally shallower profile, but the
  uncertainties are too large to draw conclusions.}
\label{fig:density}
\end{center}
\end{figure}

The number density profile can be simply calculated by using the counts from Table \ref{tab:LF}.
This is shown for magnitudes brighter than R=25 in the left panel of Fig.\ref{fig:density}. 
The solid line is the galaxy luminosity
profile with an arbitrary scaling. It is clear that the galaxy profile is a
very good representation. 

The right panel displays blue (triangles) and red (circles) clusters separately.
Considering only the three inner bins, the profile of the blue clusters might be a bit
shallower, but the uncertainty prohibits a clear statement.  The linear slope for the blue clusters is $-1.61\pm0.6$, 
for the red clusters $-1.89\pm0.3$.  The best working hypothesis is
that blue and red clusters have very similar profiles and both approximate well the galaxy  light.

\section{A spherical galaxy model}

\subsection{The brightness profile}
\label{sec:profile}
We employed the IRAF tasks {\it ellipse} and {\it bmodel} for investigating
the brightness profile of NGC 7796. The innermost 7\arcsec\ are saturated in
our image. Therefore, we used a short exposure image of NGC 7796, taken with
Gemini-South/GMOS as a preimage for a spectroscopic program
(GS-2013B-Q-83,:PI: Richtler) to fill the central part
This
image occupies the GEMINI filter g-G0325 instead of an $R$-filter, but
delivers a brightness profile, which in the overlapping region agrees
perfectly with the $R$-profile from the VIMOS image. It is a safe assumption
that within 7\arcsec\ no significant colour gradient exists, which would
prohibit such a procedure.

Since we want to use the surface brightness within a spherical model, we enforce (almost)
sphericity in the {\it ellipse}-task by fixing the ellipticity to
0.05. Objects other than the central galaxy were masked using the
``segmentation image'' produced by SExtractor (Bertin et al. 1996).

The difficulties, which arise for the profile determination at faint magnitude
levels and large radii, have to do with the fact that 
 the sky is not reached on Q2 itself. Already at a radius of 150\arcsec\, the
isophote touches the frame border and at 250\arcsec, the fraction of the
isophote covering the frame, is tiny. Therefore, the outermost value of 4625
ADUs cannot be considered to be a valid sky value.  From the other quadrants,
Q3 deviates significantly. The median value of Q1 is 4548 and of Q4 is 4557,
while it is 4138 of Q3. We therefore adopt a value of 4560 as the sky value.
An uncertainty of 0.5\%, which is perhaps difficult to
achieve, corresponds at the outermost data point to $\pm$0.3 mag
in the surface brightness.

\begin{figure}[]
\begin{center}
\includegraphics[width=0.35\textwidth]{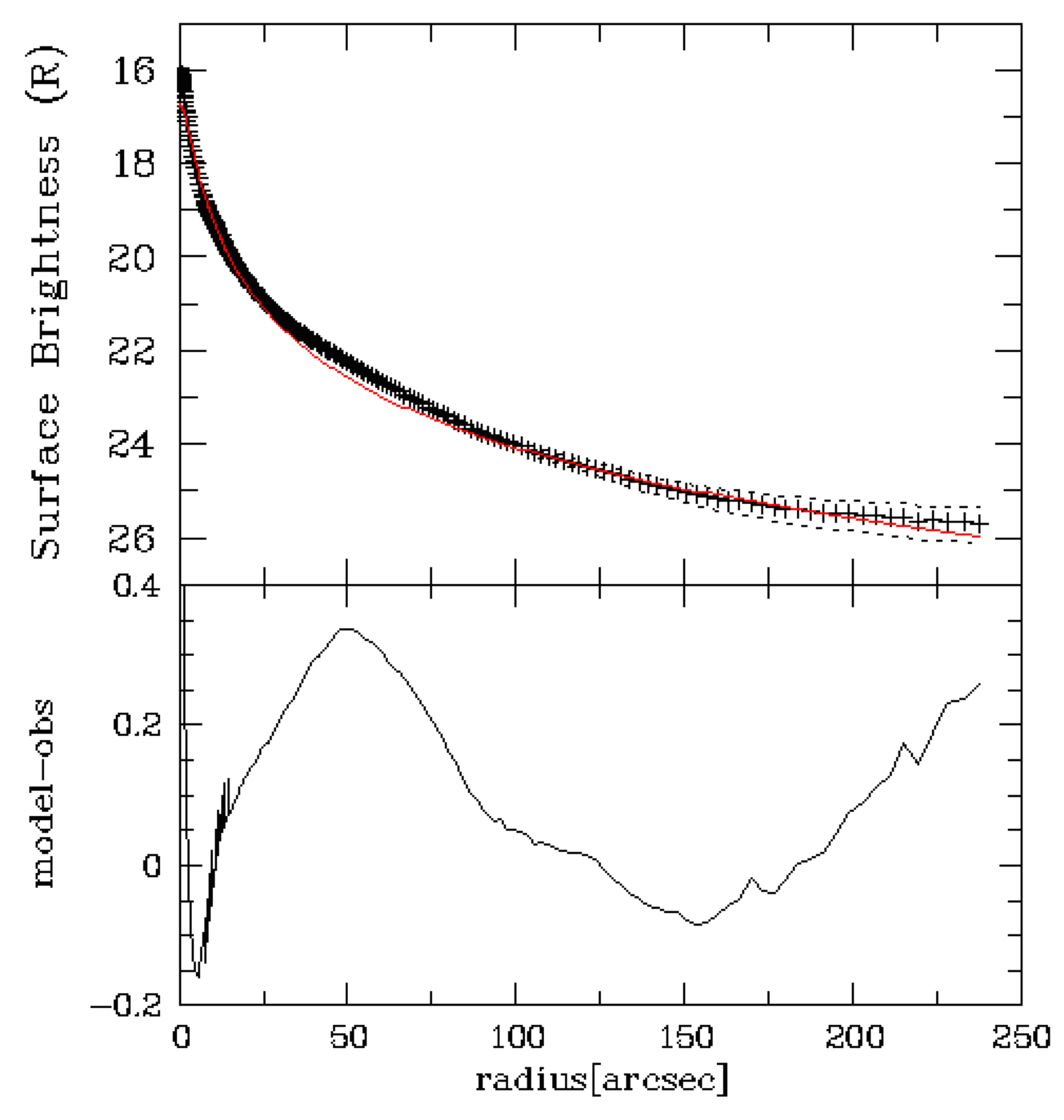}
\caption{Upper panel:  spherical surface brightness profile of NGC
  7796 ($R$-band). The red line is the beta-model (see text), which deviates
  only within the innermost few arcsec, where the measured profile is more
  cuspy than the model. The logarithmic derivative assumes the value $-2$ from
  approximately 23 arcsec outwards. Lower panel:  difference
  model$-$observations.}
\label{fig:profiles}
\end{center}
\end{figure}

Applying the calibration relations from Sec.\ref{sec:calibration} results in the
following photometric model in the R-band (mag/square arcmin) (see Fig.\ref{fig:profiles}).

\begin{equation}
\label{eq:model}
\mu(R)=-2.5\log \left(a_1 \left (1+\left( \frac{R}{R_c} \right)^2 \right)^{\alpha}\right) +8.666 
\end{equation}
with $a_1$\,=\,5.0$\times$10$^{-4}$, $R_c$\,=\,3.4\arcsec,
 $\alpha$\,=\,$-1.0$.  
 
With the data from Table \ref{tab:basicdata} this model reads in units of $L_\odot/pc^2$ 

\begin{equation}
\label{eq:modellum}
L(R)= 4.28 \times 10^3   \left(1+\left( \frac{R}{R_{pc}} \right)^2 \right)^{\alpha} 
\end{equation}
with  $R_{pc}$\,= 824 pc,
 $\alpha$\,=\,$-1.0$.  

 We used $M_{R,\odot} = 4.42$. 

\subsection{Colour map and colour profile}

\begin{figure}[]
\begin{center}
\includegraphics[width=0.4\textwidth]{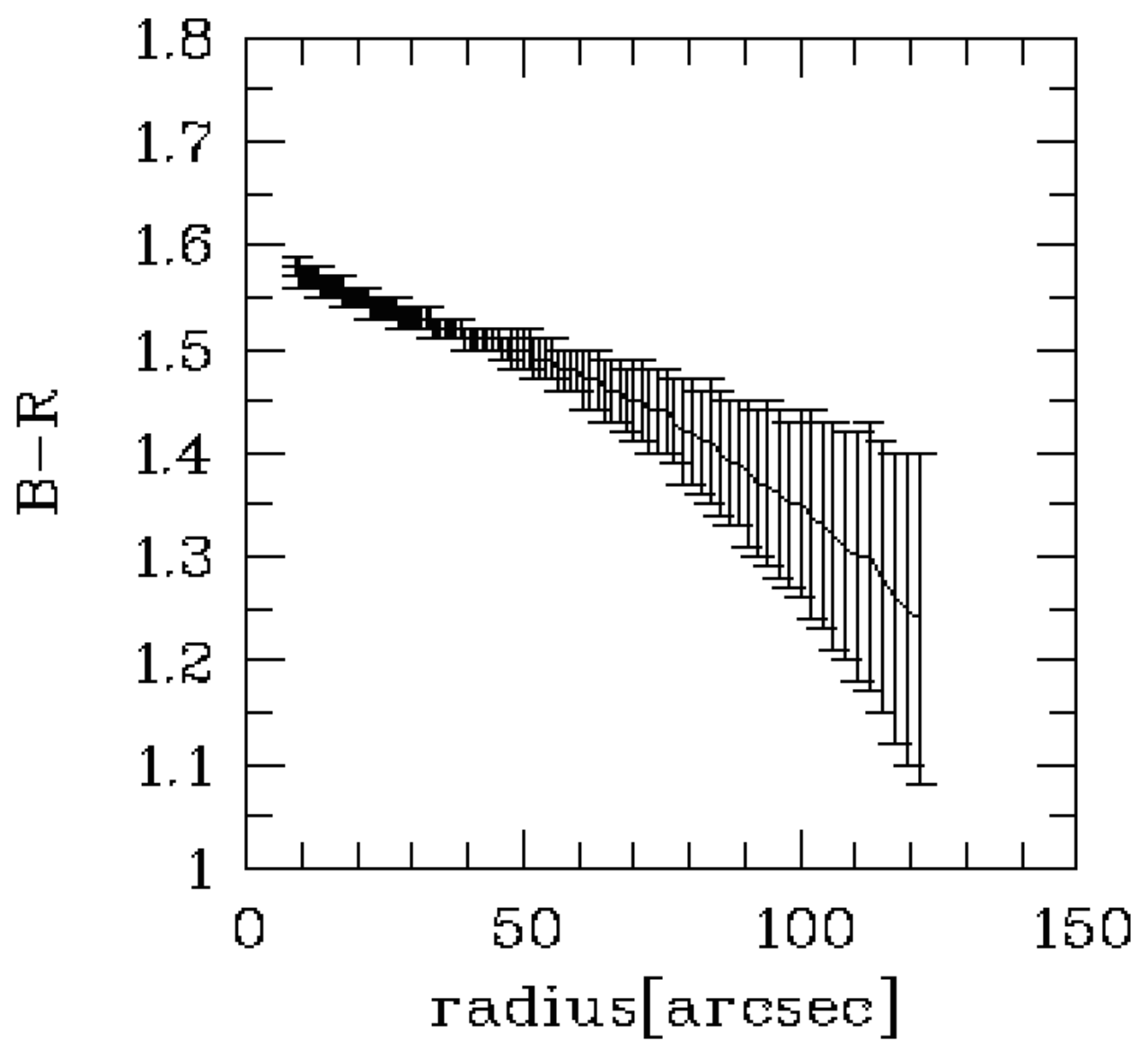}
\caption{$B-R$ colour profile for NGC 7796. The uncertainties are given by
  the {\it ellipse} task of IRAF.}
\label{fig:colourprofile}
\end{center}
\end{figure}

We constructed a colour map by applying $-2.5 \times log(B/R)+ ZP$, where B,R
denote the sky-subtracted images and ZP the zeropoint. No colour term has been
applied for the small colour range. The zeropoint has been determined by using
the aperture photometry by \citet{poulain88} by adopting a colour of
$B-R=1.59$ for the inner region outside the saturated area at a radius of
7\arcsec.
  Except for a rather smooth colour gradient, no structures are
visible, which would hint at any recent infall or merger processes.  Adopting
the models of \citet{marigo08}, the inner colour fits to a population of 12
Gyr and solar abundance.  Fig.\ref{fig:colourprofile} shows the 1D-colour
profile, which has been constructed with {\it ellipse} from the colour map.
The clear bimodality of the GCS colour distribution shows, that intermediate-age
GCs do not exist in large numbers. Therefore one concludes that 
the colour gradient is largely caused by a  metallicity gradient  and not by intermediate-age
stellar populations.

\subsection{IMACS data}
\label{sec:imacs}
IMACS data were acquired in the V-band with the idea of looking for large-scale tidal
features, indicative of merging processes. Two adjacent, overlapping,
pointings were used to produce a field of view of
$\sim27\arcmin\times40$\arcmin. Furthermore, short exposure images were
obtained to ensure that the central regions, where the VIMOS data is over
exposed, can also be analysed.

\begin{figure}[]
\begin{center}
\includegraphics[width=0.4\textwidth]{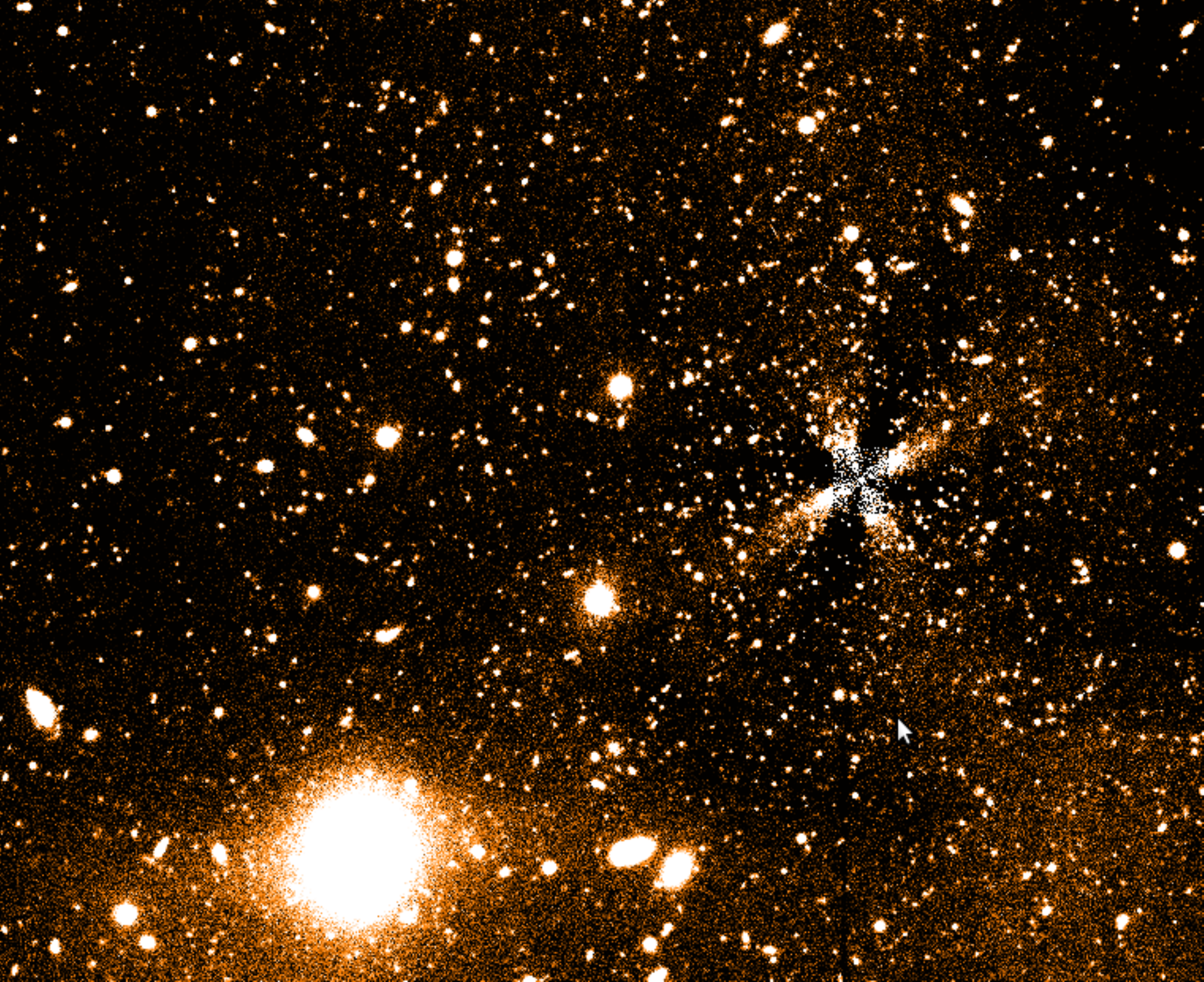}
\caption{Part of the IMACS field in the $V$-band with NGC 7796 subtracted. The
  central residuals indicate a boxy isophote symmetry, probably reminiscent of
  dry mergers. The dwarf companion is in the lower left corner. The tidal
  tails are clearly discernible. The image size is
  $6\arcmin\times6\arcmin$. North is up, east to the left.}
\label{fig:imacs}
\end{center}
\end{figure}

We removed the galaxy light using the $ellipse$ task within IRAF.
 Subtraction of the galaxy model revealed a
paucity of tidal remnants, i.e. the galaxy appears smooth and no tidal
features can be seen within these data. The central $\sim6\times6$\arcmin\ of
the galaxy-subtracted image can be seen in Fig. \ref{fig:imacs}.

The residuals from the model subtraction in the central region, within
$\sim45$\arcsec\ of the Galactic centre, show the signature of boxy isophotes
(e.g. \citealt{jedrze87}). Note that the tidal tails emanating from the dwarf companion
(see Section \ref{sec:companion}) are also obvious in the IMACS data, confirming their
reality.

\section{Dynamical remarks}

\begin{figure}[]
\begin{center}
\includegraphics[width=0.4\textwidth]{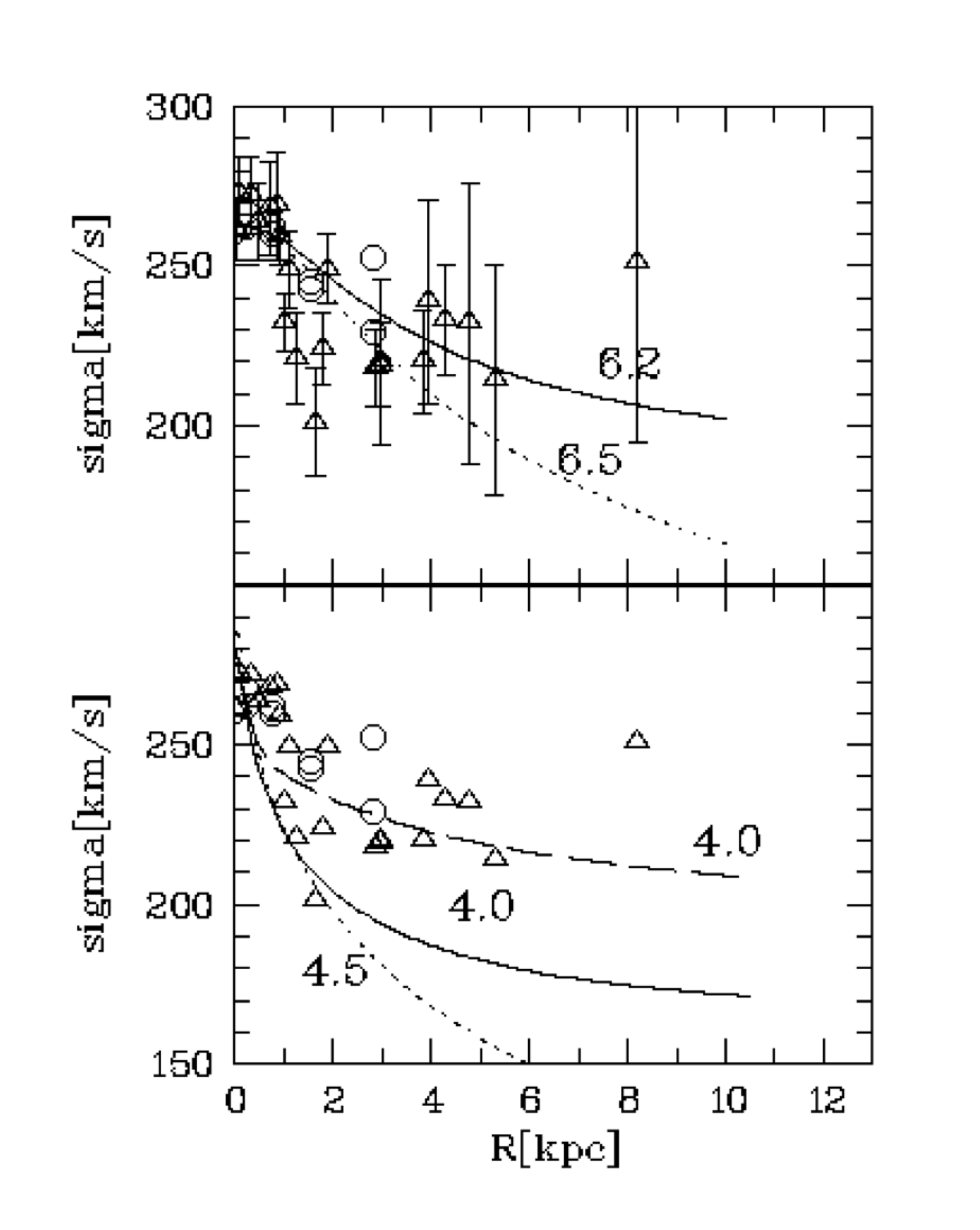}
\caption{Upper panel: the available kinematic data together with
  isotropic models. The circles are measurements by \citet{milone07}, the
  triangles by \citet{bertin94}. The central velocity dispersion of
  264\,km\,s$^{-1}$ by \citet{beuing02} is not shown. Error bars are omitted
  for readability. Plotted are galactocentric distances, irrespective of minor
  or major axis or intrinsic deviations from central symmetry. 
   The short-dashed 
  line is an isotropic model without dark matter with the stellar
  $M/L_R$-value indicated, the solid line is an isotropic MOND model.  
  Lower panel:  anisotropic models (see text).
  The solid line is a MONDian model. The
  long-dashed line is the sum of a stellar contribution with $M/L_R=4.0$ plus
  a logarithmic dark
  halo. 
   The anisotropy
  cannot be very strong in which case a MONDian halo might be a good description.}
\label{fig:models}
\end{center}
\end{figure}

The kinematical information available for NGC 7796 is relatively sparse and
not always consistent. Moreover, it is restricted to the inner 22\arcsec\ with
only one measurement at 37\arcsec. In Fig.\ref{fig:models}, we display the
velocity dispersions of \citet{bertin94} (triangles) and \citet{milone07}
(open circles).  \citet{beuing02} gives a central velocity dispersion of 264\,km\,s$^{-1}$,
which fits very well. At larger radii, the agreement is less good.  Bertin et
al. give coordinates along a long-slit, which are transformed in
Fig.\ref{fig:models} into absolute radial distances. The scatter may come from
the fact that the velocity field of NGC 7796 is not symmetric around the
centre, but the outmost point is difficult to reconcile with the inner
measurements. However, it has an uncertainty of 50\,km\,s$^{-1}$.  The data of
\citet{milone07} should have a lower signal-to-noise (S/N; smaller telescope,
shorter exposure times), but agree well in the inner region.

We now use our light model from Paper I to investigate by simple dynamical modeling,
whether a dark halo is present and what stellar M/L-values are required.
We perform spherical
Jeans modeling  as in previous work. 
 A compilation of the relevant formulae can be found
in \citet{mamon05} and \citet{schuberth10} and we refer the reader to
these papers for details. In short, we multiply the
deprojected light model with an M/L-value and calculate line-of-sight velocity
dispersions with certain assumptions concerning the anisotropy.  Regarding the
anisotropy, we use the results by \citet{hansen06} which suggest a universal
relation between the anisotropy parameter $\beta$ of the Jeans-equation and
the logarithmic slope of the density profile of the form $\beta = -0.1\times
-0.2\times slope(r)$.  We deproject our photometric model (see e.g. \citealt{schuberth10} 
for the deprojection formula) and get 
for the luminosity density j(r):
\begin{equation}
\label{eq:lumdensity}
j(r)= 2.59 \times   \left(1+\left( \frac{r}{825.4 pc} \right)^2 \right)^{-1.5}    L_\odot/pc^3 
\end{equation}

For our j(r) this means a quick rise until $\beta = 0.4$
at about 2 kpc, followed by a smooth increase until $\beta = 0.45$ at 10
kpc. This behaviour can be well approximated by formula (60) of
\citet{mamon05} so that the corresponding formulae for calculating the
line-of-sight velocity dispersion can be applied. Formula (60) then becomes
$\beta = 0.5\times r/(r+500 pc)$.
   
Fig.\ref{fig:models} shows in its upper panel the isotropic case for ``stars
only'' and the MONDian case (e.g. \citealt{milgrom12}) which has been
calculated from the Newtonian circular velocity $v_N$ by

\begin{equation}
     v_{M} = \sqrt{v_N^2(r)/2 + \sqrt{v_N^4(r)/4 + v_N^2(r)  a_0 r}} 
\end{equation}
     
\noindent where $v_{M}$ is the MONDian circular velocity and $a_0 =
1.3\times10^{-8} cm/s^2$ (see also \citealt{richtler11b}). Indicated in the
figure are the M/L-values in the $R$-band, which have been used for these
models. The requirement was to reproduce the measurements by \citet{milone07}
which here appear in our spherical model as the mean values of the minor and
major axes. The outer value of \citet{bertin94} 
cannot be reproduced within our model with any reasonable halo.

In the isotropic case, a stellar $M/L_R$ of about 6.5 is necessary to
reproduce the inner velocity dispersions. This fits well to the analysis of
the stellar population by \citet{milone07}, who found the central population
metal-rich and very old.  The SSP-models of \citet{marigo08} give $M/L_R=6.0$
for an age of 11 Gyr, a metallicity z=0.04, and a
log-normal Chabrier-IMF.  Adopting a Kroupa-IMF results in $M/L_R$=7.7. Given
the uncertainty in the distance and the fact that M/L-values are inversely
proportional to the assumed distance, the model values are in reasonable
agreement and the need for additional dark matter cannot be concluded.
However, these M/L-values are {\it global} M/L-values. As the colour gradient 
of the galaxy light shows, this global values must be lower than the central M/L-values.
We need to measure the population properties at larger radii before better models can be delevoped. 

Assuming MOND (solid line), the M/L-values do not change much. In our model, the MONDian
ghost-halo (a spherical Newtonian halo, which is added to the stellar mass and
has the MONDian circular velocities) corresponds to a mass of about
$6\times10^{10} M_\odot$ within 10 kpc.

The radially anisotropic case faces some difficulties. 
As the lower panel of
Fig.\ref{fig:models} shows, dark matter is clearly needed to match  all
measured velocity dispersions.  
But since a radial anisotropy boosts the
central dispersion values, lower stellar M/L-values are required. If dark
matter is added, these values must be even lower and cause a contradiction to
the population analysis in the centre. This is illustrated by showing a
logarithmic halo (long-dashed line), whose circular velocities are
\begin{equation} 
 v_{circ}(r) = \frac{v_0 r}{\sqrt{r_0^2+r^2}}
 \end{equation}
 
In the present case, $v_0 = 350$km/s and $r_0 = 2.5$kpc, while $M/L_R = 4.0$,
which is incompatible with the age of the central stellar population. Whether it is
compatible with the global M/L, remains to be investigated. Moreover, the
halo has a central density of about 1.1$M_\odot/pc^3$. Such a high dark matter
density is unprobable \citep{tortora09,napolitano10}. 
We conclude that at best mild radial anisotropies are
consistent with the measured velocity dispersion within our spherical model.
More remarks in Section \ref{sec:xray}.

The solid line is a MONDian model with $M/L_R =4.0$. Increasing this value  to match
the observations would result in a too high velocity dispersion in the centre. Therefore, a MONDian
behaviour of NGC 7796 seems to be possible in case of a weak radial anisotropy, but better
kinematical data are required for such an analysis.  

\section{The companion galaxy APMUKS(BJ) B235639.76-554544.9}
\label{sec:companion}
\begin{figure}[]
\begin{center}
\includegraphics[width=0.4\textwidth]{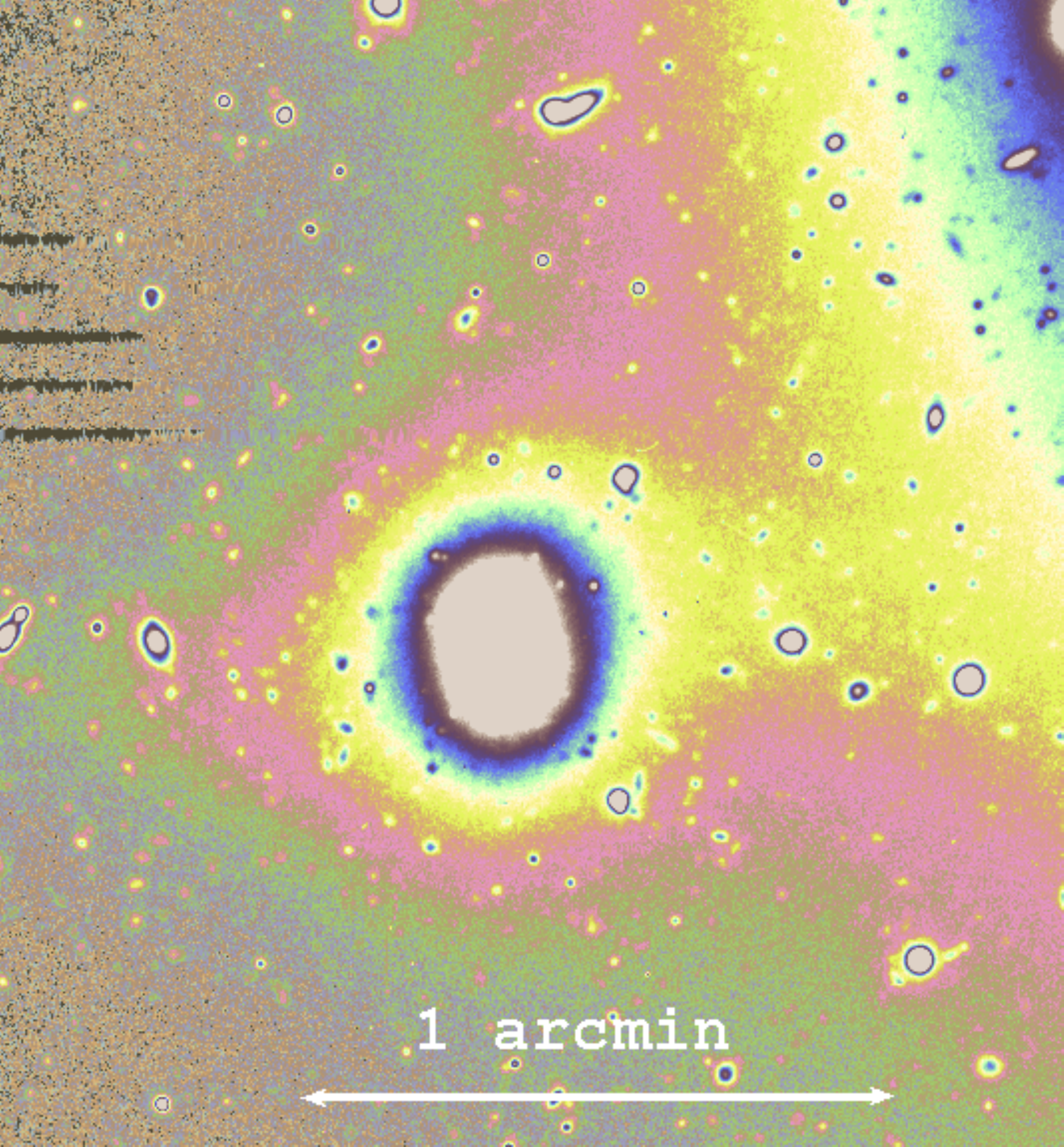}
\caption{Companion of NGC 7796 (display: skycat, random6,log). One sees the
  tidal tails, the isophote twists, and the inner very boxy isophotes. North
  is up, east to the left.}
\label{fig:tidaltails}
\end{center}
\end{figure}

\begin{figure}[h]
\begin{center}
\includegraphics[width=0.4\textwidth]{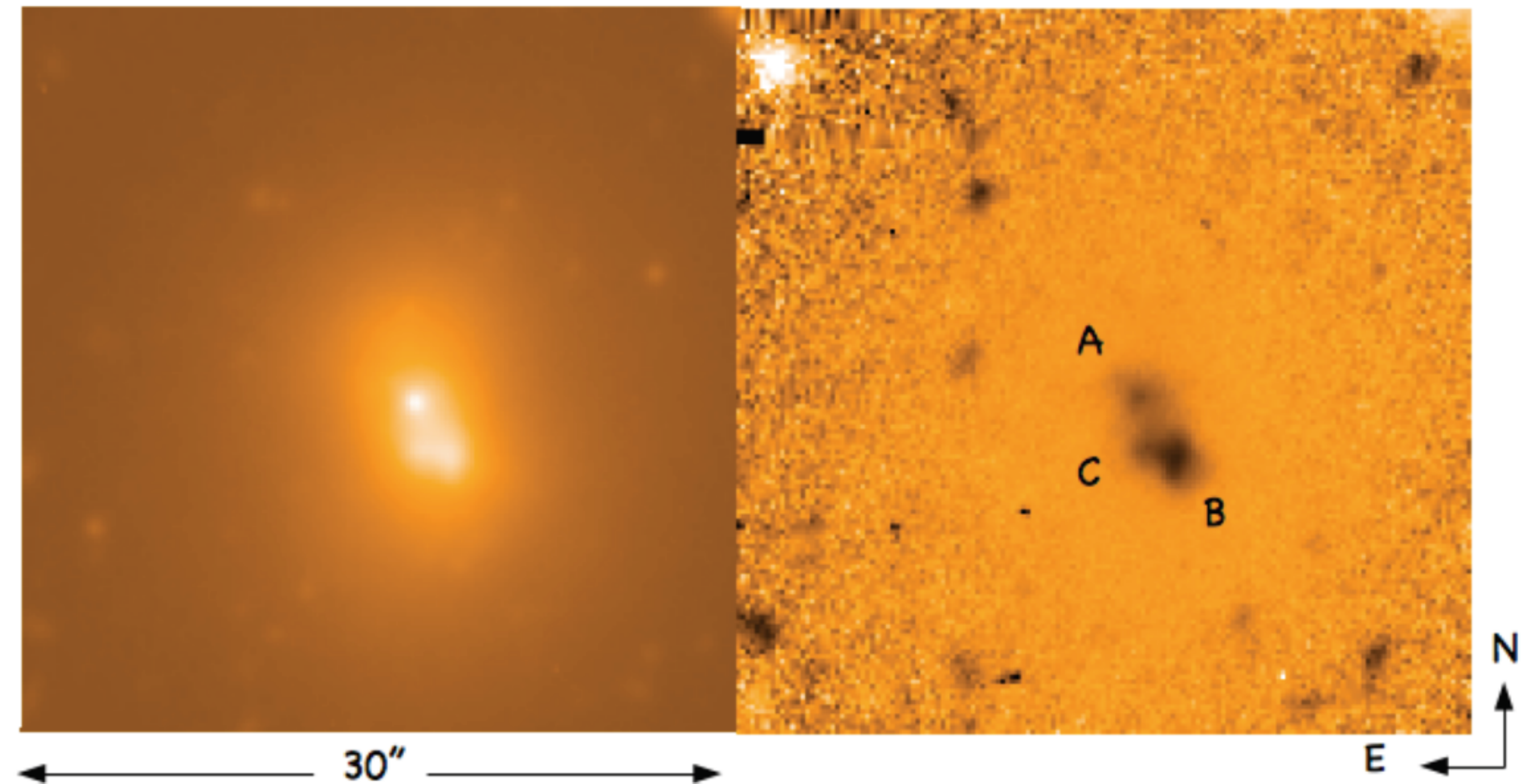}
\caption{ Left panel: central region of the dwarf companion with
  the multiple nuclei. The side length is 6.55 kpc.  Right panel: colour map where dark corresponds
  to bluer colours and bright corresponds to redder colours. See text for more
  details.}
\label{fig:companion1}
\end{center}
\end{figure}

There are some
galaxies, presumably dwarf galaxies,  around NGC 7796, but the radial velocities of these objects are
unknown. The only galaxy, where a connection with NGC 7796 is obvious, is a
dwarf galaxy 2.6\arcmin\ to the south-east. It appears in the list of
\citet{vader94} as NGC7796-1 (the NED
identification is given in the Section title).  \citet{vader94} give 
general photometric properties, classify it as a non-nucleated dwarf, but  mention a light excess. 
The properties which makes this object particularly
interesting, are: the tidal tail, the multiple nuclei, the changing of ellipticity
and position angle, and the very boxy inner isophotes, which may indicate a previous merger \citep{khochfar05}. These features are
visible in Fig.\ref{fig:tidaltails} and Fig.\ref{fig:companion1}.

The left panel of Fig.\ref{fig:companion1} shows the inner area of
$30''\times30''$, where three sources are visible. For better readability, we
identify them on the right panel with A,B,and C.  Source A is the brightest
and most compact. It gives the impression of the generic nucleus, while the
other sources are more diffuse. The projected distance between A and B is about 590 pc.
 The right panel is a colour map of the same
region. Outside the central sources, the colour is $B-R=1.25$, but for the
sources A, B and C, the bluest colours are $B-R=0.92$, $B-R=0.75$, $B-R=0.95$,
respectively. These objects also appear as slightly extended sources in the
photometric catalogue.  Table \ref{tab:nuclei} lists their magnitudes and
colours as they appear in our photometric catalogue. They may be approximate due to the
extreme crowding and the extended nature of the sources.

The metallicity is unknown, however, assuming [Fe/H]$\approx-0.5$, as is
typical of a dwarf galaxy of this brightness, the SSP models by
\citet{marigo08} provide the ages and masses (via M/L-values) presented in Table \ref{tab:nuclei}.

\begin{table}[h]
\caption{Properties of NGC7796-1 and its nuclei. Ages and masses are derived from SSP models according
  to \citet{marigo08} assuming [Fe/H]$=-0.5$. The R-magnitude of NGC7796-1 derives from the B-mag of \citet{vader94}
  and our B-R colour.}
\begin{center} 
\resizebox{0.5\textwidth}{!}{
\begin{tabular}{cccccc}
Object  &  R  &  B-R   & $ M_R$  &  age[$\times10^{9}$\,yr] &  mass[$\times 10^6$ $M_\odot$]  \\
\hline
NGC7796-1  &    15.68    &     1.25          &     -17.8       &      2.7   &  930 \\
A                     &     21.1     &     1.06          &     -12.4    &     1.7   &  4.6   \\
B                     &     21.3     &      0.65         &      -12.2    &     0.96   & 2.8  \\
C                     &     21.5    &       0.83        &       -12.0     &    1.4  &  2.8   \\
\hline
\end{tabular} }
\end{center}
\label{tab:nuclei}
\end{table}%

\begin{figure}[h]
\begin{center}g 
\includegraphics[width=0.5\textwidth]{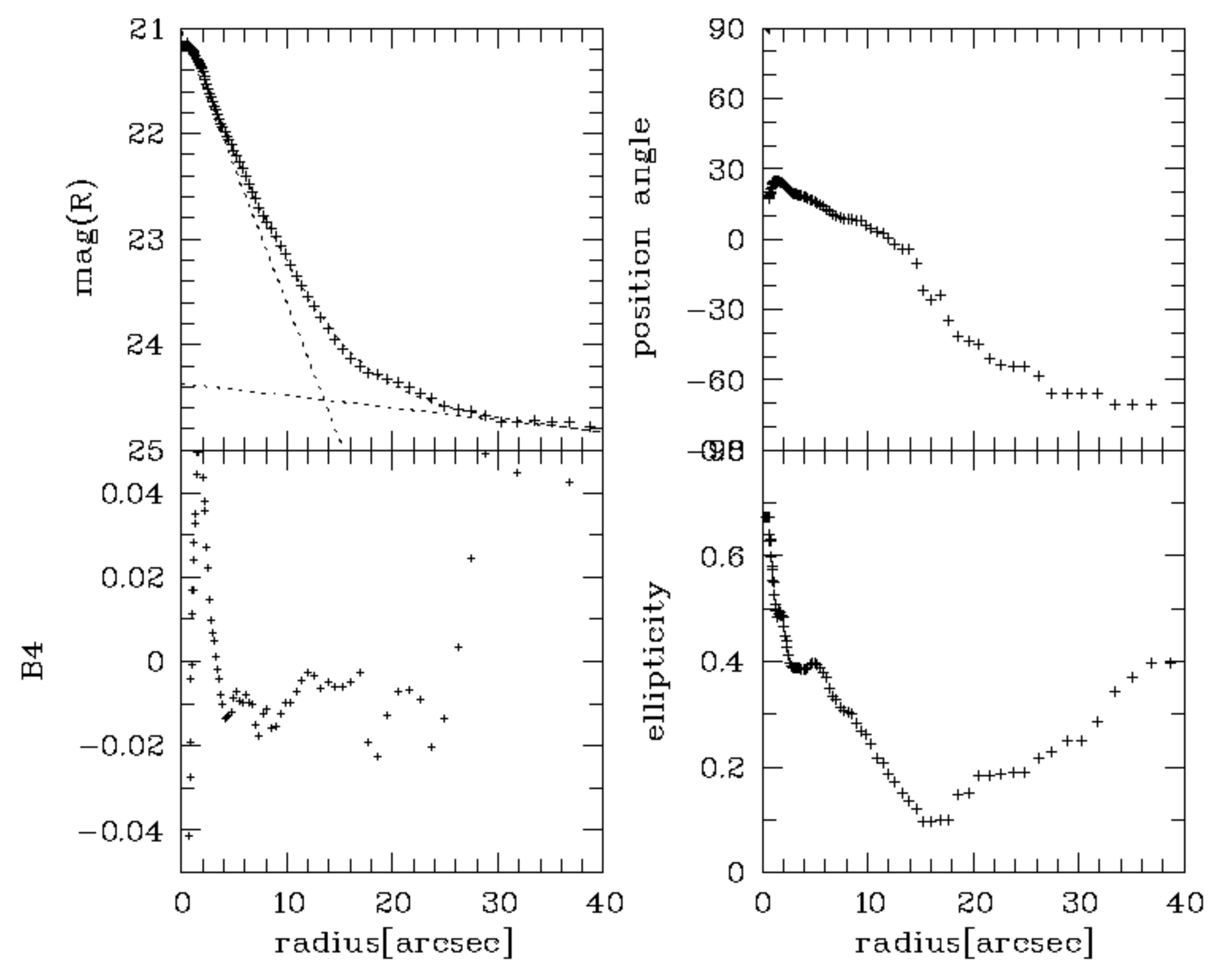}
\caption{ Upper left: surface brightness profile  of NGC7796-1, as measured by {\it ellipse}, if the position angle is allowed to vary.  The two short-dashed lines indicate the two
components of the exponential profiles.
 Lower left: The fourth harmonic cosine-term (B4) as defined by {\it ellipse}.  Negative values indicate boxy isophotes.
Upper right: Position angle of the major axis. From north to east: positive, from north to west: negative. Note that  isophotes have position angles similar 
to the axis connecting the nuclei A and B.  At larger radii, the isophotes are determined by the tidal tails and the light of NGC 7796. Lower right: ellipticity. The variation at larger
radii is due to the tidal tail and the neighbouring NGC 7796. }
\label{fig:companion2}
\end{center}
\end{figure}

We derive the light profile from an elliptical model of the galaxy. 
 Before running  the {\it ellipse}-task, we subtracted all point sources around NGC7796-1 and masked the remaining
extended sources (not the central sources). The isophote centre was fixed 1.4\arcsec\ from source A on the line connecting A and B.  We applied  {\it ellipse} varying the position angle and the ellipticity. Fig. \ref{fig:companion2} shows the results.  For radii larger than 15\arcsec\ , the surface brightness profile (upper left) is influenced by the light from NGC 7796 and from the tidal tails. 
A good representation of the light profile n the R-band is the sum of two exponentials:

\begin{equation}
\label{eq:comp_profile}
\begin{split}
 \mu(R) = -2.5*log((10^{(-0.4*21.014)}\times exp(-R/4.19)) + \\
                  10^{(-0.4*24.368)}\times exp(-R/94.5))               
 \end{split}                  
\end{equation}  
where R is in arcsec. The outer profile (see Fig.\ref{fig:companion2}) is determined by the tidal tails. 
The effective radii are for the two components $R_{eff}=7.01 \arcsec$(=1.7 kpc, inner) and $R_{eff}=158.6 \arcsec$(=38.45 kpc, outer).
The total effective radius is 32\arcsec\ (=3.05 kpc).

The inner part of the profile
is also well described by a Sersic profile with an effective radius of 10.68\arcsec\ (=2.59 kpc)
\begin{equation}
\begin{split}
 \mu(r) = 23.271+(2.1705*1.203-0.3551)*\\((R/10.683)^{(1/1.203))}-1)
 \end{split}
\end{equation}                

In nearby galaxy clusters, the early-type dwarf galaxies show a surprisingly constant effective radius, independent of brightness,
of about 950 pc \citep{smithcastelli08,misgeld08}. Fig.6 of \citet{misgeld11}  shows that NGC 7796-1 would be located at the upper limit
of the distribution and its large $R_{eff}$ might well be caused by a former merging process. 

The shifts of the position angle (upper right) and the ellipticity (lower right) are like the outer profile not  intrinsic properties of NGC7796-1.

The ellipticity passes through a minimum, because the direction towards the tidal tails and NGC7796  almost coincides with   the inner minor axis. Thus with increasing radius,
the brightness along the inner minor axis  declines slower than that along the major axis, until the isophotes  become circular and the outer major axis develops. It is an interesting
observations that the position angle of the connection of the nuclei A and B, which is 33$^\circ$, is quite similar to the isophote ellipticity also at larger radii, where the sources themselves
do not influence the isophote. Or in other words: the sources know about the shape on the larger scale, which might support the scenario that both are nuclei of the pre-merger components. 

\section{Discussion}
In the following, we comment on some particularly interesting topics and
properties of NGC 7796 and its cluster system.

\subsection{Bimodality}
The bimodal colour structure as in Fig.\ref{fig:colourhist}, which is already visible
in Fig.\ref{fig:CMDs}, although well known in GCSs for many years, has again
become a topic of intense discussion recently (for a recent review, see  e.g. \citealt{richtler13}).
The question was, whether a bimodal colour distribution also means a bimodal
metallicity distribution \citep{richtler06,yoon06,chies12}. However, the main difficulty in connecting the
colour distribution of GCs with formation scenarios is the fact that for
metal-poor clusters, some colours, in our case B-R,  becomes largely insensitive  for metallicity.
Varying the horizontal-branch morphology at a given metallicity strongly influences the 
integral colour (e.g. \citealt{chies12}) and the non-linear relation of colour and metallicity depends
on the photometric bands \citep{blakeslee12}.

\begin{figure}[]
\begin{center}
\includegraphics[width=0.4\textwidth]{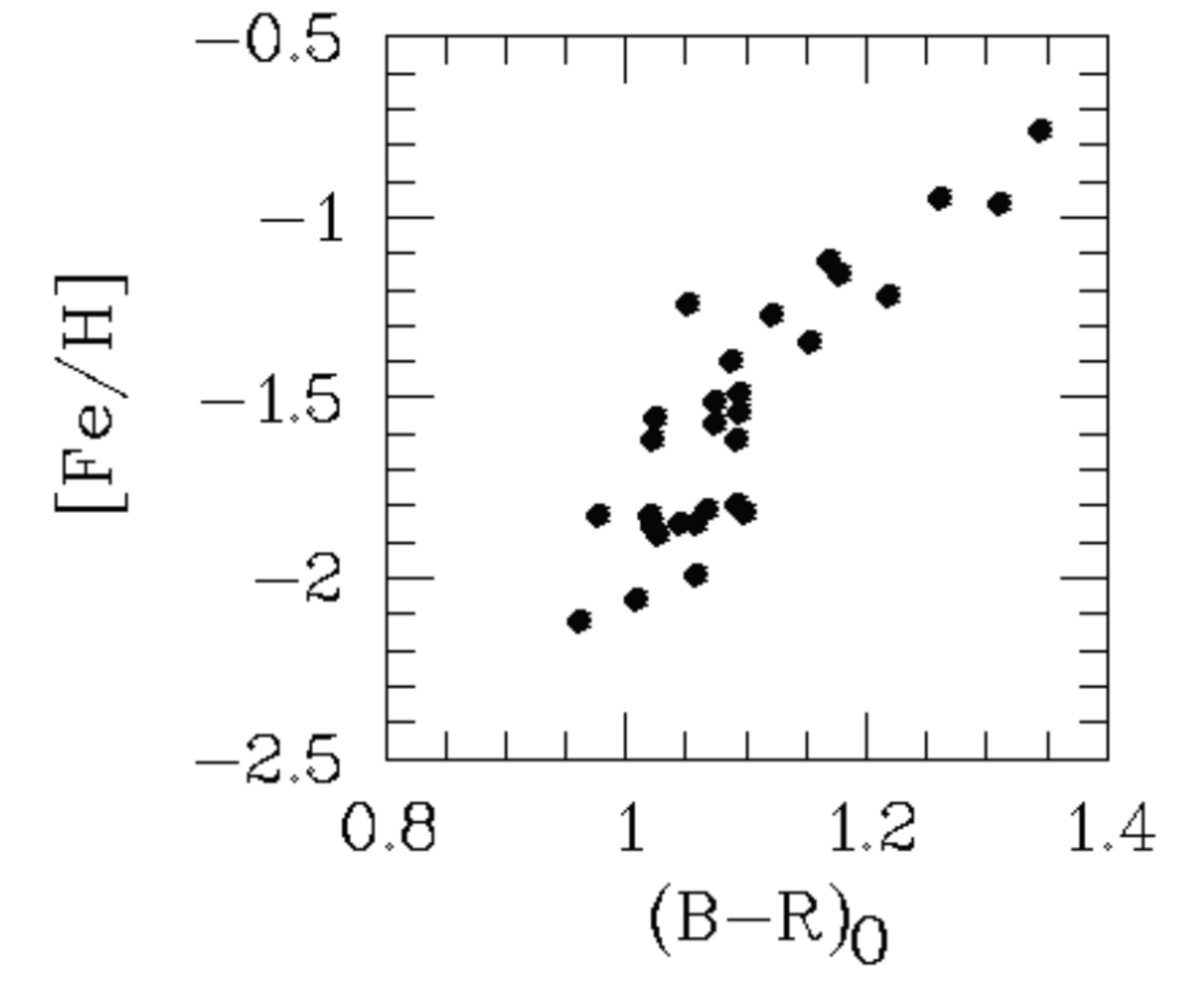}
\caption{Integral $B-R$ colours of Galactic GCs with reddenings $E(B-V)$ less
  than $0.15$ versus metallicities. Data are from \citet{harris96} (2010
  edition).  Colour is no longer a reliable indicator  for
  metallicities less than about [Fe/H]$=-1.3$}.  This is the main reason for
  the existence of a ``blue peak''.
\label{fig:galactic}
\end{center}
\end{figure}

The only objects for which we have reliable metallicities and integral
colours, are Galactic GCs.  As an argumentation aid, we plot the integral
$B-R$-colours for Galactic GCs with $E(B-V)$-values less than $0.15$, 
adopted from \citet{harris96}(2010 edition). One
reads from Fig.\ref{fig:galactic} that for metallicities less than
[Fe/H]$=-1.3$ and for colours bluer than $B-R=1.2$, the relation between
colour and metallicity practically ceases to exist. This is the reason for the
``blue peak'': a large metallicity interval is projected into a narrow colour
interval. We have no possibility to reconstruct the metallicity distribution
within the mentioned interval from broad band colours alone. The red peak, on
the other hand, consists of GCs which have been formed together with the
metal-rich bulk population of the galaxy. Thus this type of bimodality is
found generally in giant ellipticals, even if the metallicity distribution may
be not bimodal. However, \citet{usher12} present metallicities, based on the Calcium triplet
for a large sample of clusters in 11 early-type galaxies and generally found a bimodal
metallicity distribution.

Having said this, the apparent ``blue tilt'', as might be present here, hardly
indicates a strict magnitude-metallicity relation, unless the population structure of
these clusters is a strict function of metallicity, different from the
situation in the Milky Way. On the other hand, massive star clusters need
massive molecular clouds to form, most probably by the merging of star cluster
complexes.  This has been convincingly argued in the
case of W3, the most massive cluster known \citep{fellhauer05}.  
Another very
illustrative example, how massive star clusters form, is the young
globular cluster in NGC 6946  described by \citet{larsen01}. A cluster of about $5-8 \times
10^5 M\odot$ is surrounded by a plethora of less massive clusters in a presumably disk-like configuration
where  also dust is present. 
These complexes exist in starburst galaxies or at least in galaxies with a high star formation rate and  metal-poor
environments are probably forbidden.
Moreover, the accretion of dwarf galaxies  is partly responsible for the richness of globular cluster systems 
of galaxies (see \citealt{richtler13} for a review). It is thus plausible that the most massive clusters have been
donated on the average by the most massive galaxies, which have enhanced metallicities.   

\subsection{Colour profile, population synthesis, chemical evolution}

Whether age and/or metallicity is responsible for the observed colour
gradient, cannot be known from broad-band photometry. However, in old
ellipticals metallicity is supposed to affect the colour more strongly than
age. A recent survey of 33 elliptical galaxies \citep{greene13} reveals that
at 2 $R_{eff}$, the population is typically metal-poor, old, and has an
enhanced alpha-to-iron abundance, which is most easily explained by enrichment
from SNII. Whether this applies also to IEs, is not known. 


The population synthesis for NGC 7796 of \citet{milone07}, where 7 populations are
involved, results in a quite different mix of age and metallicity.  These
authors adopt three old populations (G1,G2,G3) with ages between $10-15$\,Gyr
and metallicities between solar and $-1.1$ dex. Moreover, four young
populations (Y1,Y2,Y3,RHII) with ages between $0.01-0.08$\,Gyr and
metallicities between $-0.2$ and $-0.5$ dex.  Even within their quite
restricted radial range, Milone et al. see gradients in the global metallicity
and in the alpha elements-to-iron ratio, in the sense that the centre is
alpha-overabundant ([Fe/H]$\approx0.4$) and declines to the solar value at
their outermost point along the major axis, at about 13\arcsec.  There the
old populations make up 80\% of the flux at 5870\AA, while 20\% is
provided by the young populations.
 
This is not easily brought into agreement with our colour profile. Let us
assume for simplicity only two populations, a 13 Gyr old population with
[Fe/H]$=0$ dex and a 0.025\,Gyr population with [Fe/H]$=-0.4$ dex. The model
colours are $B-R=1.67$ and $B-R=0$, respectively.  To reproduce the observed
colour at about 10\arcsec\ 
 of $B-R= 1.57$,
the young population can only contribute with 2\% in the $R$-band.
More knowledge will come from new deep and wide-field spectroscopy.

\subsection{Companion}
Also in other ``isolated'' ellipticals, we find dwarf companions with tidal
tails \citep{lane13}.  This fact  indicates that this is not a rare
event.  These dwarf galaxies may be ``late stragglers'' from
an initially much larger population which formed, in large part, NGC 7796.  If
only the central parts are enriched by SNII, the outer parts by SNIa, which
start to become active after about 2\,Gyr, one can imagine a scenario in which
a group of intermediate-mass gas-rich galaxies have fallen in at early times
and became enriched by the intermediate-age stellar population already
present.  Because the metal-poor part of the GC population is not more
extended than the galaxy light itself, these infalling galaxies can not have
been very rich in GCs themselves.

A recent paper described the first example of a merging of dwarf galaxies
around Andromeda \citep{amorisco14} as a demonstration of the hierarchical
structure of the merging process. 
Simulations of dwarf-dwarf mergers have recently been presented by \citet{cloet14} and
\cite{deason14}. 
 
We do not know for certain whether NGC7796-1
is a merger product, but its properties are very suggestive of this. In particular, its
large effective radius might indicate the case (i) of \citet{cloet14}, where a massive progenitor
grows with only a few infalling satellites. 
 Future
spectroscopic work will lend more clarity. If so, NGC7796 provides an even
more convincing example than \citet{amorisco14} of the scale-free character of
this process.
Regarding brightness and the blue nucleus, NGC 7796-1 shows some
similarity
to NGC 205 with its blue nucleus and recent star formation
\citep{monaco09}. 
\subsection{Comparison with the X-ray mass, agreements and disagreements}
\label{sec:xray}
The dynamical mass, based on X-ray-observations with Chandra, has been
evaluated by \citet{osullivan07}, who describe the X-ray luminosity profile as
a beta-profile.  Using their $\beta$-value, the scale-radius $r_c$, and the
assumption of a constant temperature, the mass is \citep{grego01}
\begin{equation}
\label{eq:xray}
M(r) =  \frac{3 \beta k T_e}{G \mu m_p} \frac{r^3}{r_c^2+r^2} 
\end{equation}
where G is the constant of gravitation, $\mu$ the molecular weight, and $m_p$
the proton mass.   Within $5R_{eff}$ (\citealt{osullivan07}
adopt $R_{eff}=25.2\arcsec$), they find an $M/L_B$-value of 10.6, compatible
with an old, metal-rich stellar population without the need for a massive dark
halo. For example, \citet{marigo08} gives $M/L_B=11.0$ for z$=0.04$ and an age of
11.2\,Gyr. The corresponding mass is $4.42\times10^{11} M_\odot$.
At small radii, they find  the resulting $M/L$ ratio to be smaller than
permitted by the stellar population. 
 O'Sullivan et al. mention the possibilities that the X-ray gas may
not be in hydrostatic equilibrium, but in a Galactic wind state, or that the
dark halo has such a large core that the increase in $M/L$ becomes apparent
only at larger radii.

Applying the above formula with $kT_e=0.53$\,keV, $\beta=0.39$,
$r_c=0.35\arcsec$, and $\mu=0.6$, we get, at a radius of 126\arcsec\ 
 or 30.54\,kpc, a mass of $6.8\times10^{11}M_\odot$ and
a (deprojected) $R$-luminosity of $6.05\times10^{10}L_{R,\odot}$, which means
$M/L_R=11.2$ and about 50\% of dark matter. The difference to the values of
O'Sullivan et al. probably has its origin in the brightness of NGC 7796, because
our X-ray masses agree well with their Fig.5. Moreover, we cannot confirm that
the inner M/L-values are lower than permitted by stellar population properties.
For example, within 10 kpc, we find $M/L_R= 5.6$.

One may also compare the X-ray mass and luminosity with the relation quoted by
\citet{kim13}: $L_{X}/10^{40} erg/s = M_{tot}/3\times10^{11}$, where $M_{tot}$ is the
total mass within 5 effective radii and $L_X$ the X-ray luminosity in the band 0.3-8 keV.
Five effective radii are 38.54 kpc and the total mass according to eq. \ref{eq:xray} within
this radius is $1.1\times10^{12} M_\odot$, where we would expect $2.2\times10^{12} M_\odot$
adopting a luminosity of $7.24\times10^{40}$ erg/s from \citet{osullivan07}. 

 The MONDian mass with a stellar $M/L_R= 6.2$ at 38.54 kpc is
$9.8\times10^{11}M_\odot$, which is in quite good agreement  with the X-ray based
mass. This M/L-value is quite high, but will lower with any 
radial anisotropy. 
 Another independent value comes from \citet{magorrian01}, who modelled the
kinematic data of \citet{bertin94} and give a circular velocity of
$371$\,km\,s$^{-1}$ at 37\arcsec. The MONDian model predicts 377\,km\,s$^{-1}$
at this radius.

One can conclude on the basis of X-rays that NGC 7796 is a promising candidate for an isolated galaxy 
with a MONDian behaviour.  The available kinematical data are not yet precise enough to
make stronger statements. We remind the case of NGC 7507, which in the sample of
\citet{krona00} was one of the most dark matter dominated galaxies and lost this status completely in the
investigation of \citet{salinas12}.  

On the other hand,  \citet{milgrom12} found good agreement with the MONDian predictions for the  isolated ellipticals 
NGC 1521 and NGC 720. \citet{memola11} investigated NGC 7052 and NGC 7785 and found disagreement with MOND.
This issue is still open and NGC 7796 might become an important cornerstone in this respect, once a thorough stellar dynamical
investigation reaching large radii, becomes available.

\section{Summary and conclusions}

We investigated the globular cluster system (GCS) of the isolated elliptical galaxy
NGC 7796 on the basis of imaging in $B$ and $R$ with VLT/VIMOS.  Moreover,
we measured the galaxy's light profile and provided possible dynamical models
based on kinematical data from the literature. Additionally, we investigated the
structure and morphology of the neighbouring dwarf galaxy NGC7796-1.\\

Our main findings/conclusions are the following:\\

-- The GCS has about 2000 members, and is  relatively
rich in comparison to other isolated ellipticals.  The specific frequency is
about 2.5$\pm$0.6. \\

-- The colour distribution of clusters resembles that of old giant ellipticals
in galaxy clusters showing a clear bimodality, whose origin is mainly
the non-linear relation between colour and metallicity. The colour distribution spans 
quite precisely the entire metallicity range, which is expected for old globular clusters. No indication of a
younger population of globular clusters has been found.\\

-- No structural abnormalities  such as shells or
ripples, common among isolated ellipticals, have been found. This matches the understanding
of NGC 7796 as an old galaxy, where signs of merging or tidal interactions have already disappeared.  \\

-- A very massive dark halo seems to be excluded. The central velocity dispersion can
be reproduced by the normal $M/L$-value for an old, metal-rich
population. However, some dark matter, for instance as required for the
baryonic Tully-Fisher relation, might be present. The available kinematical data do not permit strong conclusions 
regarding MOND,  but  the X-ray data support it as well as does
 earlier dynamical modelling.
  More certainty must be left for future
work which will measure the entire velocity field and the kinematic properties at larger radii.\\
 
-- The companion of NGC 7796 shows tidal tails, demonstrating its interaction
with NGC 7796. It also presents multiple nuclei and boxy isophotes, suggestive
of a merger. \\

NGC 7796 appears to be a genuine elliptical galaxy.
That isolated elliptical galaxies, as  listed  in catalogues,  do not belong to a class with common class
properties can be seen from different angles. The X-ray properties of the 
 \citet{memola09} sample cover a broad range of luminosities. There are old galaxies like NGC 7796, but also galaxies with more
recent star formation events and galaxies with morphological peculiarities that are indicative of 
galaxy interactions. The investigation of the dark matter properties of isolated ellipticals is still at its beginning. 
There is support  for  MOND, but also some tension there, as well as tension with the usual CDM scenario. Stellar dynamical tracers at large radii
are needed for deeper insights.

\begin{acknowledgements}
TR acknowledges  FONDECYT project Nr.\,1100620
and the BASAL Centro de Astrof\'isica y Tecnolog\'ias Afines (CATA) PFB-06/2007.
 He also acknowledges gratefully
a  visitorship at ESO/Garching,  where this paper has been completed.  RRL gratefully
acknowledges financial support from FONDECYT, project No. 3130403. This
research has made use of the NASA/IPAC Extragalactic Database (NED) which is
operated by the Jet Propulsion Laboratory, California Institute of Technology,
under contract with the National Aeronautics and Space Administration.
\end{acknowledgements}

\bibliographystyle{aa}
\bibliography{N7796_V11_arxiv.bib}

\begin{thebibliography}{73}
\expandafter\ifx\csname natexlab\endcsname\relax\def\natexlab#1{#1}\fi

\bibitem[{{Amorisco} {et~al.}(2014){Amorisco}, {Evans}, \& {van de
  Ven}}]{amorisco14}
{Amorisco}, N.~C., {Evans}, N.~W., \& {van de Ven}, G. 2014, ArXiv e-prints

\bibitem[{{Bertin} {et~al.}(1994){Bertin}, {Bertola}, {Buson}, {Danzinger},
  {Dejonghe}, {Sadler}, {Saglia}, {de Zeeuw}, \& {Zeilinger}}]{bertin94}
{Bertin}, G., {Bertola}, F., {Buson}, L.~M., {et~al.} 1994, \aap, 292, 381

\bibitem[{{Bettoni} {et~al.}(2001){Bettoni}, {Galletta}, \&
  {Prada}}]{bettoni01}
{Bettoni}, D., {Galletta}, G., \& {Prada}, F. 2001, \aap, 374, 83

\bibitem[{{Beuing} {et~al.}(2002){Beuing}, {Bender}, {Mendes de Oliveira},
  {Thomas}, \& {Maraston}}]{beuing02}
{Beuing}, J., {Bender}, R., {Mendes de Oliveira}, C., {Thomas}, D., \&
  {Maraston}, C. 2002, \aap, 395, 431

\bibitem[{{Blakeslee} {et~al.}(2012){Blakeslee}, {Cho}, {Peng}, {Ferrarese},
  {Jord{\'a}n}, \& {Martel}}]{blakeslee12}
{Blakeslee}, J.~P., {Cho}, H., {Peng}, E.~W., {et~al.} 2012, \apj, 746, 88

\bibitem[{{Buote} {et~al.}(2002){Buote}, {Jeltema}, {Canizares}, \&
  {Garmire}}]{buote02}
{Buote}, D.~A., {Jeltema}, T.~E., {Canizares}, C.~R., \& {Garmire}, G.~P. 2002,
  \apj, 577, 183

\bibitem[{{Caso} {et~al.}(2013){Caso}, {Richtler}, {Bassino}, {Salinas},
  {Lane}, \& {Romanowsky}}]{caso13}
{Caso}, J.~P., {Richtler}, T., {Bassino}, L.~P., {et~al.} 2013, \aap, 555, A56

\bibitem[{{Chies-Santos} {et~al.}(2012){Chies-Santos}, {Larsen}, {Cantiello},
  {Strader}, {Kuntschner}, {Wehner}, \& {Brodie}}]{chies12}
{Chies-Santos}, A.~L., {Larsen}, S.~S., {Cantiello}, M., {et~al.} 2012, \aap,
  539, A54

\bibitem[{{Cho} {et~al.}(2012){Cho}, {Sharples}, {Blakeslee}, {Zepf}, {Kundu},
  {Kim}, \& {Yoon}}]{cho12}
{Cho}, J., {Sharples}, R.~M., {Blakeslee}, J.~P., {et~al.} 2012, \mnras, 422,
  3591

\bibitem[{{Cloet-Osselaer} {et~al.}(2014){Cloet-Osselaer}, {De Rijcke},
  {Vandenbroucke}, {Schroyen}, {Koleva}, \& {Verbeke}}]{cloet14}
{Cloet-Osselaer}, A., {De Rijcke}, S., {Vandenbroucke}, B., {et~al.} 2014,
  ArXiv e-prints

\bibitem[{{Collobert} {et~al.}(2006){Collobert}, {Sarzi}, {Davies},
  {Kuntschner}, \& {Colless}}]{collobert06}
{Collobert}, M., {Sarzi}, M., {Davies}, R.~L., {Kuntschner}, H., \& {Colless},
  M. 2006, \mnras, 370, 1213

\bibitem[{{Deason} {et~al.}(2014){Deason}, {Wetzel}, \&
  {Garrison-Kimmel}}]{deason14}
{Deason}, A., {Wetzel}, A., \& {Garrison-Kimmel}, S. 2014, ArXiv e-prints

\bibitem[{{Dirsch} {et~al.}(2003){Dirsch}, {Richtler}, {Geisler}, {Forte},
  {Bassino}, \& {Gieren}}]{dirsch03}
{Dirsch}, B., {Richtler}, T., {Geisler}, D., {et~al.} 2003, \aj, 125, 1908

\bibitem[{{Erben} {et~al.}(2005){Erben}, {Schirmer}, {Dietrich}, {Cordes},
  {Haberzettl}, {Hetterscheidt}, {Hildebrandt}, {Schmithuesen}, {Schneider},
  {Simon}, {Deul}, {Hook}, {Kaiser}, {Radovich}, {Benoist}, {Nonino}, {Olsen},
  {Prandoni}, {Wichmann}, {Zaggia}, {Bomans}, {Dettmar}, \&
  {Miralles}}]{erben05}
{Erben}, T., {Schirmer}, M., {Dietrich}, J.~P., {et~al.} 2005, Astronomische
  Nachrichten, 326, 432

\bibitem[{{Fellhauer} \& {Kroupa}(2005)}]{fellhauer05}
{Fellhauer}, M. \& {Kroupa}, P. 2005, \mnras, 359, 223

\bibitem[{{Greene} {et~al.}(2013){Greene}, {Murphy}, {Graves}, {Gunn},
  {Raskutti}, {Comerford}, \& {Gebhardt}}]{greene13}
{Greene}, J.~E., {Murphy}, J.~D., {Graves}, G.~J., {et~al.} 2013, \apj, 776, 64

\bibitem[{{Grego} {et~al.}(2001){Grego}, {Carlstrom}, {Reese}, {Holder},
  {Holzapfel}, {Joy}, {Mohr}, \& {Patel}}]{grego01}
{Grego}, L., {Carlstrom}, J.~E., {Reese}, E.~D., {et~al.} 2001, \apj, 552, 2

\bibitem[{{Hansen} \& {Moore}(2006)}]{hansen06}
{Hansen}, S.~H. \& {Moore}, B. 2006, \na, 11, 333

\bibitem[{{Harris}(1996)}]{harris96}
{Harris}, W.~E. 1996, \aj, 112, 1487

\bibitem[{{Hau} \& {Forbes}(2006)}]{hau06}
{Hau}, G.~K.~T. \& {Forbes}, D.~A. 2006, \mnras, 371, 633

\bibitem[{{Hern{\'a}ndez-Toledo} {et~al.}(2008){Hern{\'a}ndez-Toledo},
  {V{\'a}zquez-Mata}, {Mart{\'{\i}}nez-V{\'a}zquez}, {Avila Reese},
  {M{\'e}ndez-Hern{\'a}ndez}, {Ortega-Esbr{\'{\i}}}, \&
  {N{\'u}{\~n}ez}}]{hernandez08}
{Hern{\'a}ndez-Toledo}, H.~M., {V{\'a}zquez-Mata}, J.~A.,
  {Mart{\'{\i}}nez-V{\'a}zquez}, L.~A., {et~al.} 2008, \aj, 136, 2115

\bibitem[{{Jedrzejewski}(1987)}]{jedrze87}
{Jedrzejewski}, R.~I. 1987, \mnras, 226, 747

\bibitem[{{Karachentseva}(1973)}]{karach73}
{Karachentseva}, V.~E. 1973, Astrofizicheskie Issledovaniia Izvestiya
  Spetsial'noj Astrofizicheskoj Observatorii, 8, 3

\bibitem[{{Karachentseva} {et~al.}(1997){Karachentseva}, {Lebedev}, \&
  {Shcherbanovskij}}]{karach97}
{Karachentseva}, V.~E., {Lebedev}, V.~S., \& {Shcherbanovskij}, A.~L. 1997,
  VizieR Online Data Catalog, 7082, 0

\bibitem[{{Kauffmann}(1996)}]{kauffmann96}
{Kauffmann}, G. 1996, \mnras, 281, 487

\bibitem[{{Khochfar} \& {Burkert}(2005)}]{khochfar05}
{Khochfar}, S. \& {Burkert}, A. 2005, \mnras, 359, 1379

\bibitem[{{Kim} \& {Fabbiano}(2013)}]{kim13}
{Kim}, D.-W. \& {Fabbiano}, G. 2013, \apj, 776, 116

\bibitem[{{Kronawitter} {et~al.}(2000){Kronawitter}, {Saglia}, {Gerhard}, \&
  {Bender}}]{krona00}
{Kronawitter}, A., {Saglia}, R.~P., {Gerhard}, O., \& {Bender}, R. 2000, \aaps,
  144, 53

\bibitem[{{Kuntschner} {et~al.}(2002){Kuntschner}, {Smith}, {Colless},
  {Davies}, {Kaldare}, \& {Vazdekis}}]{kuntschner02}
{Kuntschner}, H., {Smith}, R.~J., {Colless}, M., {et~al.} 2002, \mnras, 337,
  172

\bibitem[{{Landolt}(1992)}]{landolt92}
{Landolt}, A.~U. 1992, \aj, 104, 340

\bibitem[{{Lane} {et~al.}(2013){Lane}, {Salinas}, \& {Richtler}}]{lane13}
{Lane}, R.~R., {Salinas}, R., \& {Richtler}, T. 2013, \aap, 549, A148

\bibitem[{{Larsen} {et~al.}(2001){Larsen}, {Brodie}, {Huchra}, {Forbes}, \&
  {Grillmair}}]{larsen01}
{Larsen}, S.~S., {Brodie}, J.~P., {Huchra}, J.~P., {Forbes}, D.~A., \&
  {Grillmair}, C.~J. 2001, \aj, 121, 2974

\bibitem[{{Magorrian} \& {Ballantyne}(2001)}]{magorrian01}
{Magorrian}, J. \& {Ballantyne}, D. 2001, \mnras, 322, 702

\bibitem[{{Mamon} \& {{\L}okas}(2005)}]{mamon05}
{Mamon}, G.~A. \& {{\L}okas}, E.~L. 2005, \mnras, 363, 705

\bibitem[{{Marcum} {et~al.}(2004){Marcum}, {Aars}, \& {Fanelli}}]{marcum04}
{Marcum}, P.~M., {Aars}, C.~E., \& {Fanelli}, M.~N. 2004, \aj, 127, 3213

\bibitem[{{Marigo} {et~al.}(2008){Marigo}, {Girardi}, {Bressan}, {Groenewegen},
  {Silva}, \& {Granato}}]{marigo08}
{Marigo}, P., {Girardi}, L., {Bressan}, A., {et~al.} 2008, \aap, 482, 883

\bibitem[{{Memola} {et~al.}(2011){Memola}, {Salucci}, \&
  {Babi{\'c}}}]{memola11}
{Memola}, E., {Salucci}, P., \& {Babi{\'c}}, A. 2011, \aap, 534, A50

\bibitem[{{Memola} {et~al.}(2009){Memola}, {Trinchieri}, {Wolter}, {Focardi},
  \& {Kelm}}]{memola09}
{Memola}, E., {Trinchieri}, G., {Wolter}, A., {Focardi}, P., \& {Kelm}, B.
  2009, \aap, 497, 359

\bibitem[{{M{\'e}ndez} {et~al.}(2009){M{\'e}ndez}, {Teodorescu}, {Kudritzki},
  \& {Burkert}}]{mendez09}
{M{\'e}ndez}, R.~H., {Teodorescu}, A.~M., {Kudritzki}, R.-P., \& {Burkert}, A.
  2009, \apj, 691, 228

\bibitem[{{Milgrom}(2012)}]{milgrom12}
{Milgrom}, M. 2012, Physical Review Letters, 109, 131101

\bibitem[{{Milone} {et~al.}(2007){Milone}, {Rickes}, \& {Pastoriza}}]{milone07}
{Milone}, A.~D.~C., {Rickes}, M.~G., \& {Pastoriza}, M.~G. 2007, \aap, 469, 89

\bibitem[{{Misgeld} \& {Hilker}(2011)}]{misgeld11}
{Misgeld}, I. \& {Hilker}, M. 2011, \mnras, 414, 3699

\bibitem[{{Misgeld} {et~al.}(2008){Misgeld}, {Mieske}, \& {Hilker}}]{misgeld08}
{Misgeld}, I., {Mieske}, S., \& {Hilker}, M. 2008, \aap, 486, 697

\bibitem[{{Monaco} {et~al.}(2009){Monaco}, {Saviane}, {Perina}, {Bellazzini},
  {Buzzoni}, {Federici}, {Fusi Pecci}, \& {Galleti}}]{monaco09}
{Monaco}, L., {Saviane}, I., {Perina}, S., {et~al.} 2009, \aap, 502, L9

\bibitem[{{Mulchaey} \& {Zabludoff}(1999)}]{mulchaey99}
{Mulchaey}, J.~S. \& {Zabludoff}, A.~I. 1999, \apj, 514, 133

\bibitem[{{Napolitano} {et~al.}(2010){Napolitano}, {Romanowsky}, \&
  {Tortora}}]{napolitano10}
{Napolitano}, N.~R., {Romanowsky}, A.~J., \& {Tortora}, C. 2010, \mnras, 405,
  2351

\bibitem[{{Niemi} {et~al.}(2010){Niemi}, {Hein{\"a}m{\"a}ki}, {Nurmi}, \&
  {Saar}}]{niemi10}
{Niemi}, S.-M., {Hein{\"a}m{\"a}ki}, P., {Nurmi}, P., \& {Saar}, E. 2010,
  \mnras, 405, 477

\bibitem[{{O'Sullivan} {et~al.}(2007){O'Sullivan}, {Sanderson}, \&
  {Ponman}}]{osullivan07}
{O'Sullivan}, E., {Sanderson}, A.~J.~R., \& {Ponman}, T.~J. 2007, \mnras, 380,
  1409

\bibitem[{{Poulain}(1988)}]{poulain88}
{Poulain}, P. 1988, \aaps, 72, 215

\bibitem[{{Prugniel} \& {Heraudeau}(1998)}]{prugniel98}
{Prugniel}, P. \& {Heraudeau}, P. 1998, \aaps, 128, 299

\bibitem[{{Reda} {et~al.}(2004){Reda}, {Forbes}, {Beasley}, {O'Sullivan}, \&
  {Goudfrooij}}]{reda04}
{Reda}, F.~M., {Forbes}, D.~A., {Beasley}, M.~A., {O'Sullivan}, E.~J., \&
  {Goudfrooij}, P. 2004, \mnras, 354, 851

\bibitem[{{Rejkuba}(2012)}]{reijkuba12}
{Rejkuba}, M. 2012, \apss, 341, 195

\bibitem[{{Richtler}(2003)}]{richtler03}
{Richtler}, T. 2003, in Lecture Notes in Physics, Berlin Springer Verlag, Vol.
  635, Stellar Candles for the Extragalactic Distance Scale, ed. D.~{Alloin} \&
  W.~{Gieren}, 281--305

\bibitem[{{Richtler}(2006)}]{richtler06}
{Richtler}, T. 2006, Bulletin of the Astronomical Society of India, 34, 83

\bibitem[{{Richtler}(2013)}]{richtler13}
{Richtler}, T. 2013, in Astronomical Society of the Pacific Conference Series,
  Vol. 470, 370 Years of Astronomy in Utrecht, ed. G.~{Pugliese}, A.~{de
  Koter}, \& M.~{Wijburg}, 327

\bibitem[{{Richtler} {et~al.}(2011){Richtler}, {Famaey}, {Gentile}, \&
  {Schuberth}}]{richtler11b}
{Richtler}, T., {Famaey}, B., {Gentile}, G., \& {Schuberth}, Y. 2011, \aap,
  531, A100

\bibitem[{{Salinas} {et~al.}(2012){Salinas}, {Richtler}, {Bassino},
  {Romanowsky}, \& {Schuberth}}]{salinas12}
{Salinas}, R., {Richtler}, T., {Bassino}, L.~P., {Romanowsky}, A.~J., \&
  {Schuberth}, Y. 2012, \aap, 538, A87

\bibitem[{{Schirmer}(2013)}]{schirmer13}
{Schirmer}, M. 2013, \apjs, 209, 21

\bibitem[{{Schlegel} {et~al.}(1998){Schlegel}, {Finkbeiner}, \&
  {Davis}}]{schlegel98}
{Schlegel}, D.~J., {Finkbeiner}, D.~P., \& {Davis}, M. 1998, \apj, 500, 525

\bibitem[{{Schuberth} {et~al.}(2010){Schuberth}, {Richtler}, {Hilker},
  {Dirsch}, {Bassino}, {Romanowsky}, \& {Infante}}]{schuberth10}
{Schuberth}, Y., {Richtler}, T., {Hilker}, M., {et~al.} 2010, \aap, 513, A52+

\bibitem[{{Smith} {et~al.}(2004){Smith}, {Mart{\'{\i}}nez}, \&
  {Graham}}]{smith04}
{Smith}, R.~M., {Mart{\'{\i}}nez}, V.~J., \& {Graham}, M.~J. 2004, \apj, 617,
  1017

\bibitem[{{Smith Castelli} {et~al.}(2008){Smith Castelli}, {Bassino},
  {Richtler}, {Cellone}, {Aruta}, \& {Infante}}]{smithcastelli08}
{Smith Castelli}, A.~V., {Bassino}, L.~P., {Richtler}, T., {et~al.} 2008,
  \mnras, 386, 2311

\bibitem[{{Spitler} {et~al.}(2008){Spitler}, {Forbes}, {Strader}, {Brodie}, \&
  {Gallagher}}]{spitler08}
{Spitler}, L.~R., {Forbes}, D.~A., {Strader}, J., {Brodie}, J.~P., \&
  {Gallagher}, J.~S. 2008, \mnras, 385, 361

\bibitem[{{Tal} {et~al.}(2009){Tal}, {van Dokkum}, {Nelan}, \&
  {Bezanson}}]{tal09}
{Tal}, T., {van Dokkum}, P.~G., {Nelan}, J., \& {Bezanson}, R. 2009, \aj, 138,
  1417

\bibitem[{{Thomas} {et~al.}(2005){Thomas}, {Maraston}, {Bender}, \& {Mendes de
  Oliveira}}]{thomas05}
{Thomas}, D., {Maraston}, C., {Bender}, R., \& {Mendes de Oliveira}, C. 2005,
  \apj, 621, 673

\bibitem[{{Tonry} {et~al.}(2001){Tonry}, {Dressler}, {Blakeslee}, {Ajhar},
  {Fletcher}, {Luppino}, {Metzger}, \& {Moore}}]{tonry01}
{Tonry}, J.~L., {Dressler}, A., {Blakeslee}, J.~P., {et~al.} 2001, \apj, 546,
  681

\bibitem[{{Tortora} {et~al.}(2009){Tortora}, {Napolitano}, {Romanowsky},
  {Capaccioli}, \& {Covone}}]{tortora09}
{Tortora}, C., {Napolitano}, N.~R., {Romanowsky}, A.~J., {Capaccioli}, M., \&
  {Covone}, G. 2009, \mnras, 396, 1132

\bibitem[{{Turner} \& {Gott}(1976)}]{turner76}
{Turner}, E.~L. \& {Gott}, III, J.~R. 1976, \apjs, 32, 409

\bibitem[{{Usher} {et~al.}(2012){Usher}, {Forbes}, {Brodie}, {Foster},
  {Spitler}, {Arnold}, {Romanowsky}, {Strader}, \& {Pota}}]{usher12}
{Usher}, C., {Forbes}, D.~A., {Brodie}, J.~P., {et~al.} 2012, \mnras, 426, 1475

\bibitem[{{Vader} \& {Chaboyer}(1994)}]{vader94}
{Vader}, J.~P. \& {Chaboyer}, B. 1994, \aj, 108, 1209

\bibitem[{{van Dokkum} {et~al.}(2010){van Dokkum}, {Whitaker}, {Brammer},
  {Franx}, {Kriek}, {Labb{\'e}}, {Marchesini}, {Quadri}, {Bezanson},
  {Illingworth}, {Muzzin}, {Rudnick}, {Tal}, \& {Wake}}]{vandokkum10}
{van Dokkum}, P.~G., {Whitaker}, K.~E., {Brammer}, G., {et~al.} 2010, \apj,
  709, 1018

\bibitem[{{Villegas} {et~al.}(2010){Villegas}, {Jord{\'a}n}, {Peng},
  {Blakeslee}, {C{\^o}t{\'e}}, {Ferrarese}, {Kissler-Patig}, {Mei}, {Infante},
  {Tonry}, \& {West}}]{villegas10}
{Villegas}, D., {Jord{\'a}n}, A., {Peng}, E.~W., {et~al.} 2010, \apj, 717, 603

\bibitem[{{Yoon} {et~al.}(2006){Yoon}, {Yi}, \& {Lee}}]{yoon06}
{Yoon}, S.-J., {Yi}, S.~K., \& {Lee}, Y.-W. 2006, Science, 311, 1129

\end{thebibliography}

\end{document}